\documentclass[prd,notitlepage,superscriptaddress]{revtex4-1}

\usepackage{physics}
\usepackage{amsmath}
\usepackage{amssymb}  
\usepackage{graphicx}  
\usepackage{verbatim} 
\usepackage{color}    
\usepackage{subfigure}  
\usepackage{hyperref}  
\usepackage{mathrsfs}
\usepackage{float}

\newcommand{\Eotvos}{E\"{o}tv\"{o}s}

\begin{document}


\title{Generalized Analysis of the E\"{o}tv\"{o}s Experiment}

\author{M. J. Mueterthies}
\affiliation{Department of Physics  and Astronomy, Purdue University, West Lafayette, IN 47907, USA}
\author{E. Fischbach}
\email[Corresponding author:\,]{ephraim@purdue.edu}
\affiliation{Department of Physics  and Astronomy, Purdue University, West Lafayette, IN 47907, USA}
\author{V. E. Barnes}
\affiliation{Department of Physics  and Astronomy, Purdue University, West Lafayette, IN 47907, USA}
\author{J. Bertaux}
\affiliation{Department of Physics  and Astronomy, Purdue University, West Lafayette, IN 47907, USA}
\author{D. E. Krause}
\affiliation{Physics Department, Wabash College, Crawfordsville, IN 47933, USA}
\affiliation{Department of Physics  and Astronomy, Purdue University, West Lafayette, IN 47907, USA}
\author{A. Longman}
\affiliation{Department of Physics  and Astronomy, Purdue University, West Lafayette, IN 47907, USA}

\date{\today}

\date{\today}

\begin{abstract}
We present a generalized phenomenological formalism for analyzing the original \Eotvos\  experiment in the presence of gravity and a generic ``5th force.''  To date no evidence for a 5th force has emerged since its presence was suggested by a 1986 reanalysis of the 1922 publication coauthored by \Eotvos, Pek\`ar, and Fekete (EPF).  However, our generalized analysis introduces new mechanisms capable in principle of accounting for the EPF data, while at the same time avoiding detection by most recent experiments carried out to date.  As an example, some of these mechanisms raise the possibility that the EPF signal could have arisen from an unexpected direction if it originated from the motion of the Earth through a medium.
\end{abstract}


\maketitle

\section{Introduction}
\label{sec:Introduction}

The results of the well-known \Eotvos\ experiment  were published in 1922 \cite{EPF} by \Eotvos\, Pek\`ar and Fekete (EPF), following the passing of \Eotvos\ in 1919. The EPF experiment, which was a test of what is now referred to as the weak equivalence principle (WEP), quoted limits on possible WEP violations which were approximately 300 times more sensitive than those from an earlier experiment by Bessel (cited in \cite{EPF}). However, a 1986 reanalysis of the EPF results \cite{Fischbach1986} revealed a pattern in the EPF data which was consistent with the presence of a new interaction coupling to baryon number $B$, the so-called ``5th force'' \cite{Fischbach1986}. This force was attributed to the virtual exchange of massive vector bosons coupling  with fundamental ``charge'' $f$ to baryon number $B$, giving rise to a two-body potential of  the form
\begin{equation}
V_{12}(r) = f^{2}\frac{B_{1}B_{2}}{r}e^{-r/\lambda},
\label{Yukawa}
\end{equation}
where $\lambda = \hbar/m_{b} c$ determines the range of the interaction in terms of the boson mass $m_{b}$. Notwithstanding the seemingly compelling evidence for a 5th force that emerged from Ref. \cite{Fischbach1986}, a force with the characteristics suggested by Ref.~\cite{Fischbach1986} (a force of intermediate range with roughly gravitational strength) does not appear to be present in nature. (For reviews of the many experiments conducted since 1986, see Refs.~\cite{Metrologia,FischbachBook,Adelberger2009,Wagner2012,Murata2015}).

If we believe that EPF carried out their experiment correctly, and that the subsequent analysis in Ref. \cite{Fischbach1986} of the EPF data is also correct, then the failure of the community to detect a 5th force represents a puzzle which has recently been termed the ``\Eotvos\ Paradox'' \cite{Fischbach2019}. Specifically, this paradox arises from the mutual incompatibility of the following three statements, each of which appears to be correct \cite{Fischbach2019}:

\begin{enumerate}
\item There is consensus in the community that there were no obvious flaws in the EPF experiment or in the published paper \cite{EPF}.
\item There is additional consensus that the analysis of the EPF data reported in Ref. \cite{Fischbach1986} is also correct, along with its suggestion of a new ``5th force'' in nature.
\item There is no credible experimental evidence for a new force with the characteristics presented in Ref.~\cite{Fischbach1986}.
\end{enumerate}

Not surprisingly, most attempts to resolve this paradox have focused on the correctness of the EPF experiment. Considerable attention has been devoted to the possibility that some overlooked systematic influences, such as gravity gradients (e.g., Ref.~\cite{Toth2019}), could be responsible for the EPF data. To date, no compelling mechanism has been advanced which accounts for the EPF data in terms of classical environmental effects such as temperature, pressure, or humidity variations, or gravity gradients, etc. (See Ref.~\cite{Fischbach1988} for an extensive discussion and references and Ref.~\cite{Volgyesi2019} describing an attempt to repeat the EPF experiment with modern technology.) This is not entirely surprising given that the pattern in the EPF data noted in Ref. \cite{Fischbach1986} depends on a non-classical quantity, the baryon-to-mass ratio of each sample ($B/\mu$ in the notation of Ref. \cite{Fischbach1986}). Further support for the correctness of the EPF experiment comes from a recently discovered handwritten partial draft by \Eotvos\ (autograph) of what would be eventually become the published EPF paper \cite{Kilenyi2019}.  As discussed in Ref.~\cite{Fischbach2019}, this autograph, and various exchanges between \Eotvos\ and Einstein, lend further support to the inference that the EPF experiment was in fact done correctly.

The object of the present paper is to develop a more general phenomenological framework for analyzing the original torsion balance experiment of EPF, and the more modern torsion balance experiments in the presence of both gravity and a possible 5th force.   Although the focus of the present paper will be on torsion balance experiments, such as  those by the  E\"{o}t-Wash collaboration \cite{Eotwash1987,Eotwash1988,Eotwash1989},
 the formalism developed here can be extended to apply to other searches for composition-dependent forces.

The outline of our paper is as follows:  In Sec.~\ref{section:Methodology} we develop the general formalism for combining a composition-independent gravitational interaction with a general composition-dependent interaction arising from a coupling to baryon number $B$ that lead to a torque on a torsion balance.  In Sec.~\ref{section:Eotvos} we apply this formalism to the EPF experiment, taking care to include couplings which were omitted in the original EPF analysis \cite{EPF}.  In Sec.~\ref{section:Eot-Wash}, the formalism is then applied to the E\"{o}t-Wash (E-W) experiment of Stubbs, et al. \cite{Eotwash1987}.  In Sec.~\ref{sec:Implications}, we demonstrate that EPF and E-W experiments are sensitive to different 5th force signals, and in following Sec.~\ref{sec:Examples}, we illustrate this using  a toy scalar-vector 5th force model.
 We also consider a qualitatively different scenario to account for the EPF results arising from forces with a non-zero curl.  As we note, such forces could arise from a number of sources, including a ``magnetic'' 5th force, or from a dark matter medium.  What is significant about such forces in the present context is that they can appear to arise from unexpected directions, and hence could be rejected as spurious.  Since the sensitivity of the E-W experiment (and others) depends to a great extent on knowing the direction from which the presumed 5th force is emanating (such as a nearby mountain),   most experiments are optimized to seek a particular signal.  Unlike the EPF experiment, more recent force experiments have been designed to set limits on new forces described by potentials similar to Eq.~(\ref{Yukawa}).  We will suggest, however, that these experiments may not be as sensitive as the EPF experiment to other types of forces.

\section{General Force acting on a Torsion Balance}
\label{section:Methodology}

In order to determine whether a force exists that could have been detected in the \Eotvos\ experiment, but not in later experiments such as E\"ot-Wash, we consider here the response of a mass $m$ on a torsion balance to a general baryon-number-dependent force $\vec{F}_{5}$, which we will write in the form of a Taylor expansion about the position of the pivot point $\vec{R}$.  Assuming the Einstein summation convention, the $i$th component of $\vec{F}_{5}$, $F_{5i}$, will be written as
\begin{equation}
\label{e:rawfifthforce}
 F_{5i} = \xi G m q \overline{q}_{\rm source}\left[\mathcal{F}_{5i}(\vec{R})+\mathcal{D}_{5ij}(\vec{R})r_{j}+\cdots\right],
\end{equation}
where $\xi$ is a dimensionless constant, $G$ is the Newtonian gravitational constant,  $q = B/\mu$ is the baryon number-to-mass ratio  of the sample, $\overline{q}_{\rm source}$ is the average baryon-to-mass number ratio of the  source,   and $\vec{r}$ is the position of the sample relative to the pivot point.  In order for $q$ and $\overline{q}_{\rm source}$ to be dimensionless, $\mu \equiv m/m_{\rm H}$, where $m_{\rm H} = m(^{1}$H$_{1})$ is the mass of atomic hydrogen.   If $\vec{F}_{5}$ arises from a potential $V_{5} = \xi G m q \overline{q}_{\rm source}\mathcal{V}_{5}$, then $\mathcal{F}_{5i}$ is the gradient of $\mathcal{V}_{5}$, while $\mathcal{D}_{5ij}$ is the gradient of $\mathcal{F}_{5i}$: $\mathcal{F}_{5i} = -\partial_{i}\mathcal{V}_{5}$; $\mathcal{D}_{5ij} = \partial_{j}\mathcal{F}_{5i}=-\partial_{i}\partial_{j}\mathcal{V}_{5}$, and $\partial_{i} = \partial / \partial R_{i}$.
To simplify the notation, we introduce
\begin{subequations}
\begin{eqnarray}
 f_{5i} &  \equiv & \xi G \overline{q}_{\rm source}\mathcal{F}_{5i}, \label{f4i} \\
 d_{5ij} & \equiv &  \xi G \overline{q}_{\rm source}\mathcal{D}_{5ij}. \label{d5ij}
 \end{eqnarray}
 \end{subequations}
 Since both the sources and test masses in torsion balance experiments are extended objects, the 5th force charge will be allowed to vary with position, and hence we replace the mass with the differential mass: $m q\rightarrow q(\vec{r})\mathrm{d}m(\vec{r})$. The force on the differential element $\mathrm{d}m(\vec{r})$ is then given by
\begin{equation}
\label{e:fifthfor_q}
\mathrm{d}F_{5i} = q(\vec{r})\mathrm{d}m(\vec{r})\left(f_{5i}+d_{5ij}r_{j}+\cdots\right).
\end{equation}
In addition to the 5th force, the total force $\vec{F}_{\rm tot}$ will also include the gravitational force $\vec{F}_{\rm gravity}$ as well as the centripetal force $\vec{F}_{\rm cent}$ due to the rotation of the Earth. The $i$th component of the total force then becomes
\begin{equation}
F_{{\rm tot},i} = F_{{\rm grav},i}+F_{{\rm cent},i}+m\overline{q}f_{5i}+m\overline{q}\overline{\ell}_{j}d_{5ij}\text{,}
\end{equation}
where $m = \int \mathrm{d}m$, $\overline{q}= \int q(\vec{r})\,\mathrm{d}m(\vec{r})/m,$ and $\overline{\ell}_{j}= \int q(\vec{r})r_{j}\,\mathrm{d}m(\vec{r})/m\overline{q},$ and we have dropped terms higher order in $r_{i}$.

\begin{figure}[t]
  \centering
  \includegraphics[width=0.8\columnwidth]{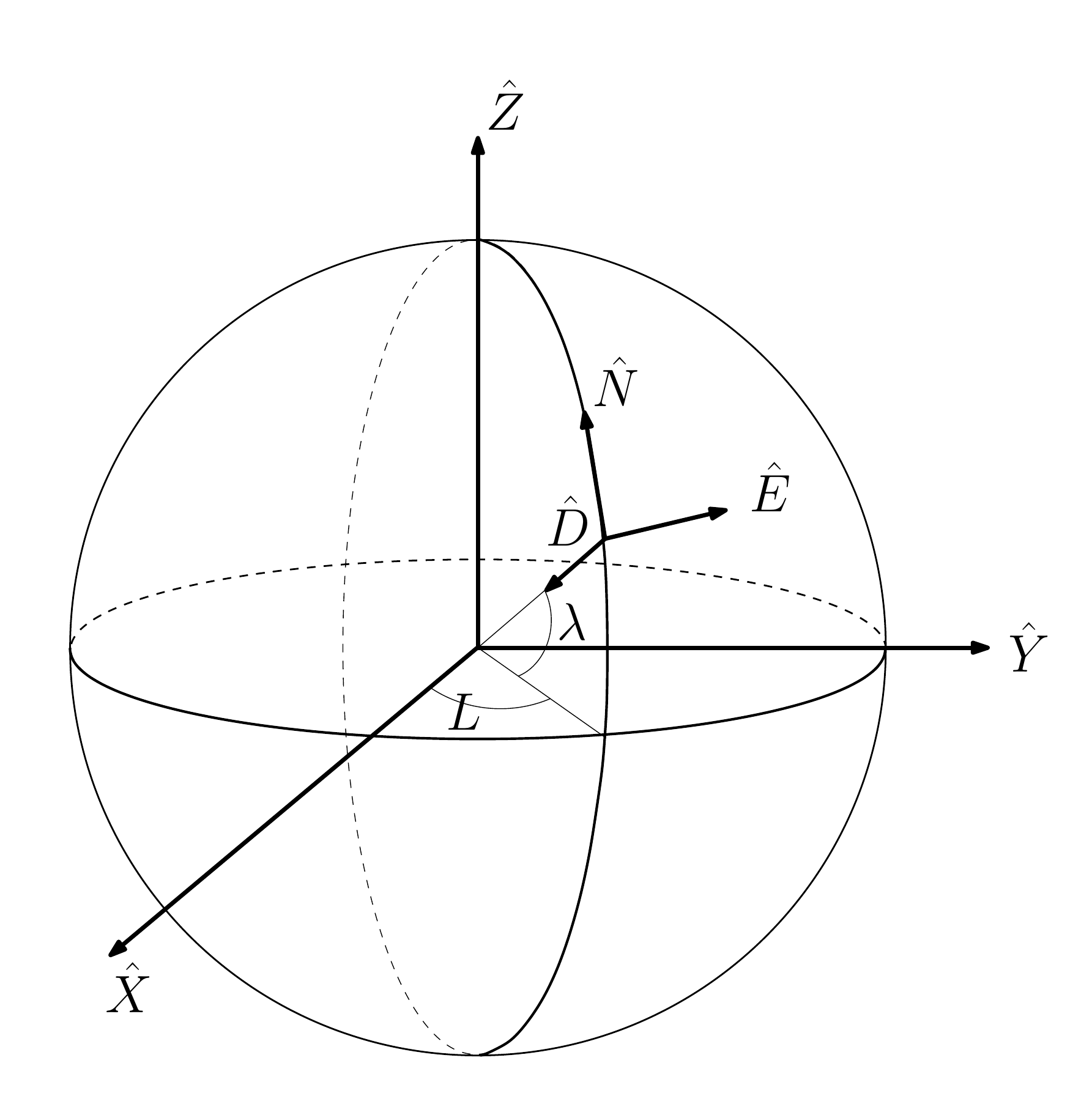}
  \caption{Representation of the NED reference frame.}
  \label{fi:EarthFrame}
\end{figure}

\begin{figure}[t]
  \centering
  \includegraphics[width=.9\columnwidth]{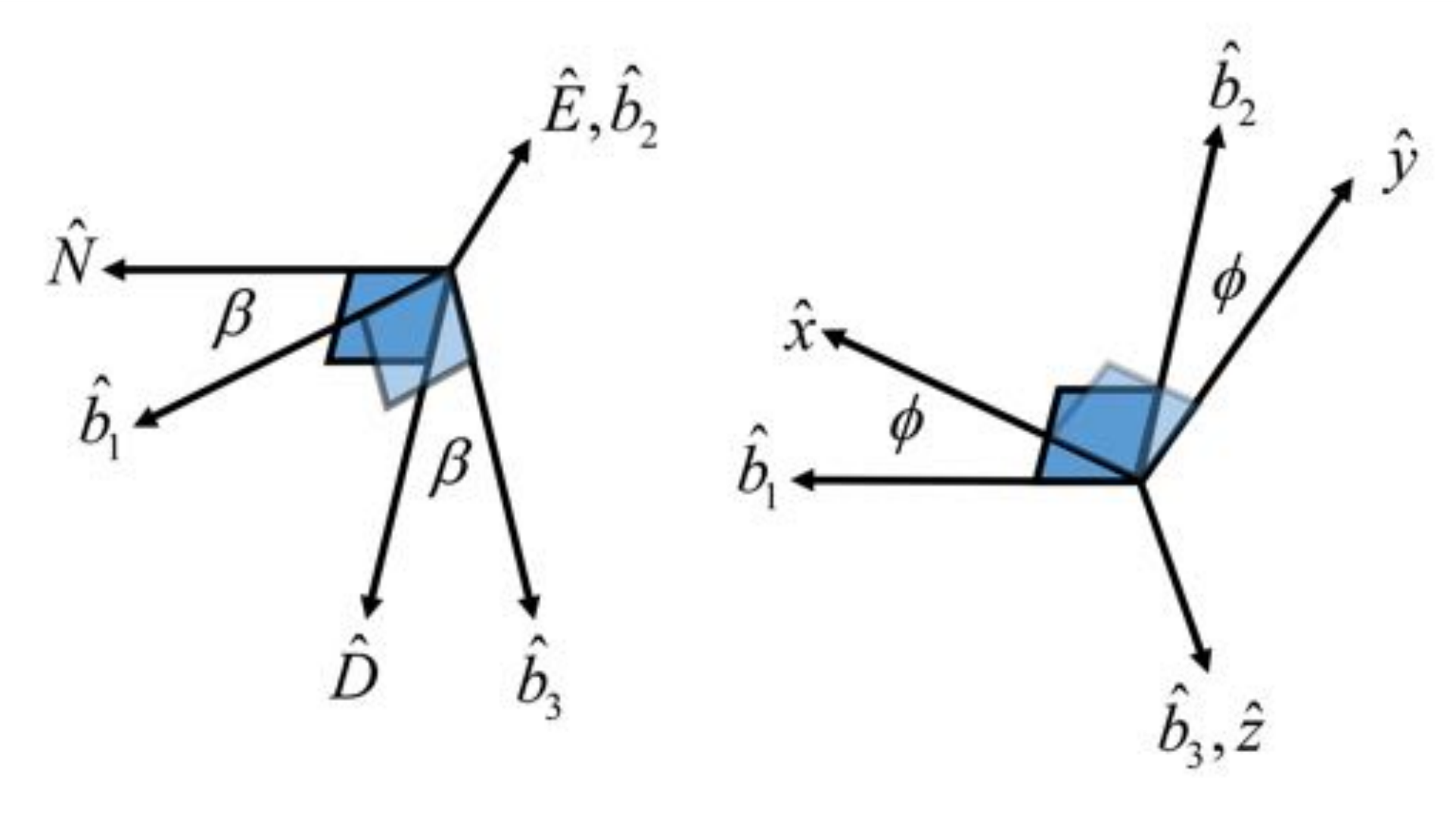}
  \caption{The transformations from the NED frame to the body frame.}
  \label{fi:NEDBody}
\end{figure}

To introduce the reference frames relevant to characterize our torsion balance system, we begin with a local North-East-Down (NED) frame as shown in Fig.~\ref{fi:EarthFrame}. The forces under consideration are defined in this frame, where the axis of the torsion balance is aligned with the local vertical, found by hanging a plumb bob. The rotation axis $\hat{b}_{3}$ of the torsion balance is then given by 
\begin{equation}
\label{e:b3}
\hat{b}_{3} = \frac{\vec{F}_{\rm tot}}{|\vec{F}_{\rm tot}|},
\end{equation}
while the $\hat{b}_{1}$-axis is along the torsion balance beam when the intrinsic twist in the fiber is zero. Since it is convenient for this axis to be roughly towards North ($\hat{N}$), this can be achieved if we define the $\hat{b}_{2}$-axis, the axis perpendicular to the torsion balance, as
\begin{equation}
\label{e:b2}
\hat{b}_{2} = \frac{\hat{b}_{3}\times\hat{N}}{|\hat{b}_{3}\times\hat{N}|},
\end{equation}
and then let $\hat{b}_{1}$  be given by
\begin{equation}
\label{e:b1}
\hat{b}_{1} = \hat{b}_{2}\times\hat{b}_{3}.
\end{equation}
The $\hat{b}$ frame is space-fixed in the zero-twist orientation of the beam and makes an angle $\beta$ with the NED frame as shown in Fig.~\ref{fi:NEDBody}).   If $\vec{F}_{\rm tot}$ were directed downward, $\beta = 0$ and the NED and $\hat{b}$ frames would coincide.

 We also introduce the body fixed $x$-$y$-$z$ frame which is a rotation of the $\hat{b}$ frame by the angle $\theta$,
\begin{subequations}
\begin{eqnarray}
\hat{N} & = & \hat{x}c_{\phi}c_{\beta}-\hat{y}s_{\phi}c_{\beta}-\hat{z} s_{\beta},\\
\hat{E} & = & \hat{x}s_{\phi}+\hat{y}c_{\phi}, \\
\hat{D} & = & \hat{x}c_{\phi}s_{\beta}-\hat{y}s_{\phi}s_{\beta}+\hat{z} c_{\beta}.
\end{eqnarray}
\end{subequations}
Here and throughout the rest of the text, we use the notation
\begin{subequations}
\begin{eqnarray}
c_{\phi} & \equiv & \cos\phi, \\
s_{\phi} & \equiv & \sin\phi,
\end{eqnarray}
\end{subequations}
and similar expressions for angles $\beta$, etc.   The frame relations are also depicted in Figure \ref{fi:NEDBody}.

To find the torque on the torsion balance, we use the Lagrangian formulation
\begin{equation}
\label{e:EulLag}
\frac{\mathrm{d}}{\mathrm{d}t}\left(\frac{\partial\mathcal{L}}{\partial \dot{\phi}}\right)-\frac{\partial\mathcal{L}}{\partial \phi} = Q_{5\phi}\text{,}
\end{equation}
where $Q_{5\phi}$ is the generalized 5th force given by,
\begin{eqnarray}
\nonumber \mathrm{d}Q_{5\phi} & = & \frac{\partial r_{i}}{\partial \phi}\cdot \mathrm{d}\mathcal{F}_{5i}\\
\nonumber & = &q(\vec{r})\mathrm{d}m(\vec{r})\left(f_{5i}\frac{\partial r_{i}}{\partial \phi}+d_{5ij}r_{j}\frac{\partial r_{i}}{\partial \phi}\right)\\
& = &q(\vec{r})\rho(\vec{r})\mathrm{d}V\left(f_{5i}\frac{\partial r_{i}}{\partial \phi}+d_{5ij}r_{j}\frac{\partial r_{i}}{\partial \phi}\right)\text{.}
\end{eqnarray}
We note that $\vec{r}$ here, and in Eq.~(\ref{e:rawfifthforce}), is the position of the particle in the NED frame, which can be written as
\begin{eqnarray}
\nonumber \vec{r} & = & x\hat{x}+y\hat{y}+z\hat{z}\\
\nonumber & = & \left(xc_{\phi}c_{\beta}-ys_{\phi}c_{\beta}-z s_{\beta}\right)\hat{N}+\left(xs_{\phi}+yc_{\phi}\right)\hat{E}\\
&&+\left(xc_{\phi}s_{\beta}-ys_{\phi}s_{\beta}+z c_{\beta}\right)\hat{D}\text{.}
\end{eqnarray}
It is straightforward to show that the angular derivative of $\vec{r}$ is
\begin{equation}
\frac{\partial\vec{r}}{\partial \phi} = \left(-xs_{\phi}c_{\beta}-yc_{\phi}c_{\beta}\right)\hat{N}+\left(xc_{\phi}-ys_{\phi}\right)\hat{E}+\left(-xs_{\phi}s_{\beta}-yc_{\phi}s_{\beta}\right)\hat{D}\text{.}
\end{equation}

\section{Application to the EPF Experiment}
\label{section:Eotvos}

\begin{table}[t]
  \caption{Notation correspondence between EPF (Ref.~\cite{EPF}) and this work.}
  \begin{center}
    \begin{tabular}{ccl}
	\hline      
      \bf This work& \bf EPF&  \bf Description \\
      \hline
       $I$& $K$& Moment of Inertia\\
       $V_{\rm grav}$ & $-U$& Gravitational Potential Energy\\
       $\delta$& $L$& Observation scale length \\
       $L$& $\ell_{a}$& Torsion Balance length \\
       $m$& $M_{a}$& Sample mass \\
       $\beta$& $\epsilon$& Plumb bob angle \\
       $\Delta\phi$& $\Delta\alpha$ &Intrinsic twist of the wire\\
      $\sigma$ & $m=n_{E}-n_{W}$& East-West observation difference\\
	\hline
\end{tabular}
  \end{center}
  \label{ta:NotationDifferences}
\end{table}

\begin{figure}[t]
  \centering
  \includegraphics[width=0.8\columnwidth]{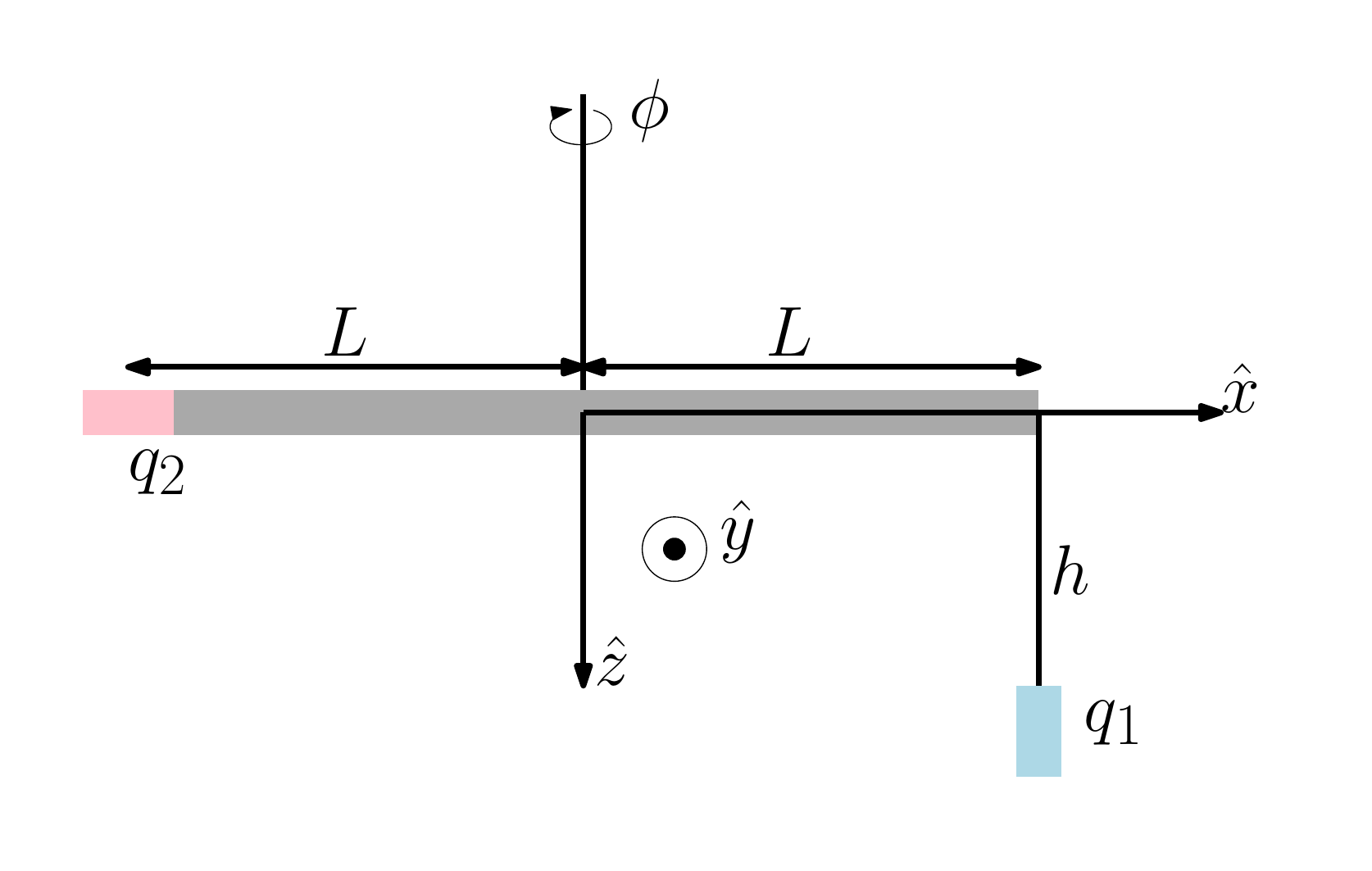}
  \caption{Side view of the E\"otv\"os Experiment.}
  \label{fi:EotvosSide}
\end{figure}

To be more consistent with current practices, we have modified the notation of Ref.~\cite{EPF} as shown in Table~\ref{ta:NotationDifferences}. The EPF experiment compared two masses, as shown in Fig.~\ref{fi:EotvosSide}. Mass 1 was located along the balance towards North and hung a distance $h$ below the balance. Mass 2 was located along the balance towards the South and was located at the same height as the balance. 

\begin{figure}[t]
  \centering
  \includegraphics[width=0.8\columnwidth]{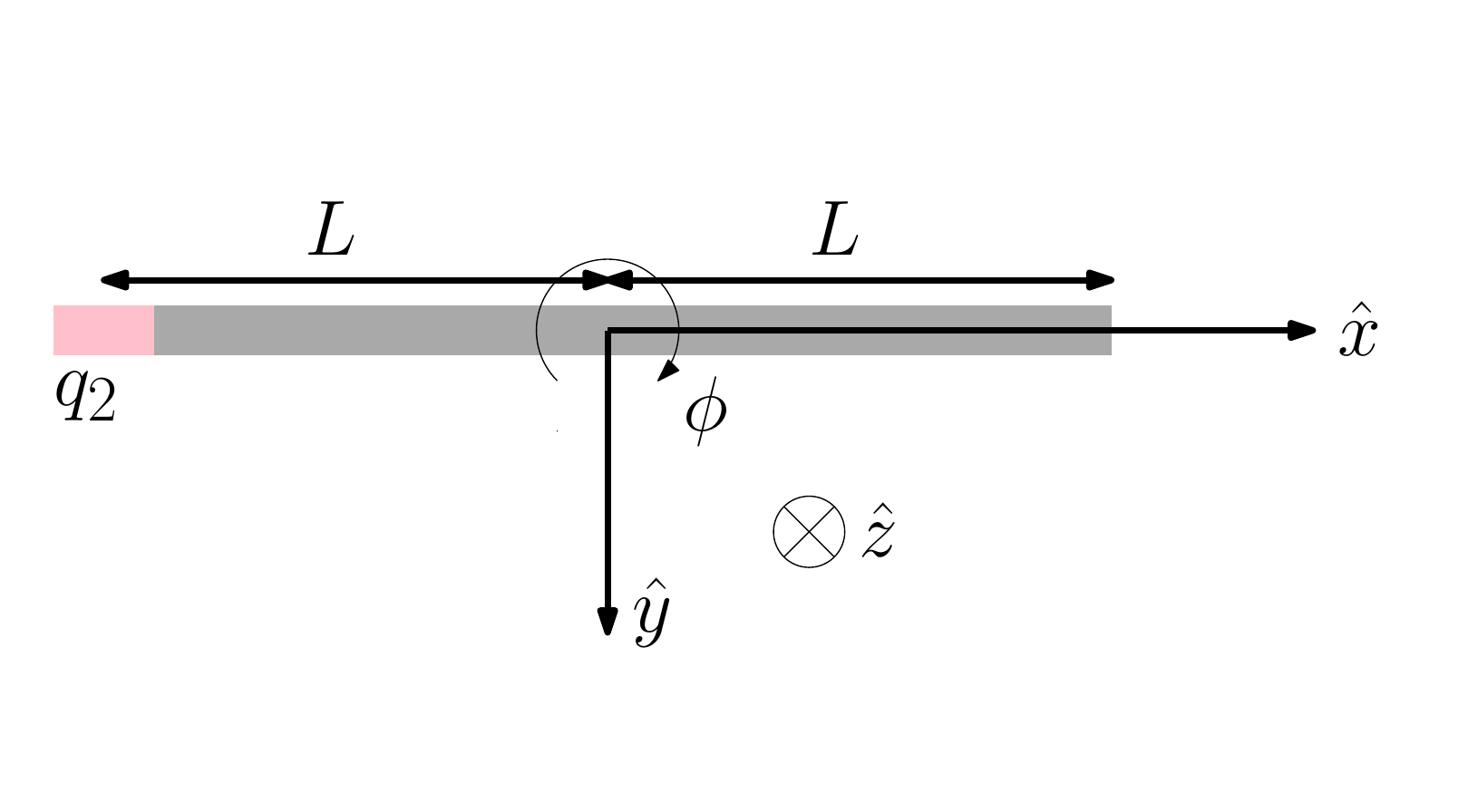}
  \caption{Top view of the E\"otv\"os Experiment.}
  \label{fi:EotvosTop}
\end{figure}

The product of the 5th force charge and the mass density of the EPF samples can be written as,
\begin{equation}
\label{e:qm_EPF}
q\left(\vec{r}\right)\rho\left(\vec{r}\right) = q_{1}\delta\left(x-L\right)\delta\left(y\right)\delta\left(z-h\right) + q_{2}\delta\left(x+L\right)\delta\left(y\right)\delta\left(z\right)\text{.}
\end{equation}
Using Eq.~(\ref{e:qm_EPF}), the torque on the fiber can then be expressed in the form
\begin{eqnarray}
\label{e:Tfifth}
\nonumber T_{5} & = & -(q_{1}-q_{2})mL c_{\beta}s_{\phi}f_{5x}
 + (q_{1}-q_{2})mL c_{\phi}f_{5y}
 - (q_{1}-q_{2})mL s_{\beta}s_{\phi}f_{5z}\\
 %
\nonumber && -(q_{1}+q_{2})\frac{I}{2}c_{\beta}^2c_{\phi}s_{\phi}d_{5xx}+q_{1}mL h s_{\beta}c_{\beta}s_{\phi}d_{5xx}\\
%
\nonumber && +(q_{1}+q_{2})\frac{I}{2}c_{\beta}c_{\phi}^{2}d_{5yx}-q_{1}mL h s_{\beta}c_{\phi}d_{5yx}\\
%
\nonumber && -(q_{1}+q_{2})\frac{I}{2}c_{\beta}s_{\beta}c_{\phi}s_{\phi}d_{5zx}+q_{1}mL h s_{\beta}^{2}s_{\phi}d_{5zx}\\
%
\nonumber && -(q_{1}+q_{2})\frac{I}{2}c_{\beta}s_{\phi}^{2}d_{5xy}
%
 +(q_{1}+q_{2})\frac{I}{2}c_{\phi}s_{\phi}d_{5yy}
%
 -(q_{1}+q_{2})\frac{I}{2}s_{\beta}s_{\phi}^{2}d_{5zy}\\
 %
\nonumber && -(q_{1}+q_{2})\frac{I}{2}c_{\beta}s_{\beta}c_{\phi}s_{\phi}d_{5xz}-q_{1}mL h c_{\beta}^{2}s_{\phi}d_{5xz}\\
%
\nonumber && +(q_{1}+q_{2})\frac{I}{2}s_{\beta}c_{\phi}^{2}d_{5yz}+q_{1}mL h c_{\beta}c_{\phi}d_{5yz}\\
%
&& -(q_{1}+q_{2})\frac{I}{2}s_{\beta}^{2}c_{\phi}s_{\phi}d_{5zz}-q_{1}mL h s_{\beta}c_{\beta}s_{\phi}d_{5zz}\text{.}
\end{eqnarray}

We also need the gravitational and centrifugal force vectors, which can be determined in two ways. First, we could form the Lagrangian using the gravitational potential energy and the effective potential energy for the centrifugal force. Then we expand the Lagrangian to second order in $\vec{r}$,  and finally apply Eq.~(\ref{e:EulLag}) to find the torque. Alternatively, we can use the results from Eq.~(\ref{e:Tfifth}) to quickly deduce the remaining terms, as Eq.~(\ref{e:Tfifth}) was written generically. We simply let $q_{1}=q_{2}=1$, $f_{i} = -\partial_{i}V_{\rm grav}+F_{{\rm cent},i}/m$, and $d_{ij} = -\partial_{i}\partial_{j}V_{\rm grav}$, where $V_{\rm grav}$ is the gravitational potential and $F_{{\rm cent},i}$ is the $i$th component of the centrifugal force. The Newtonian contribution to the torque is then given,
\begin{eqnarray}
\nonumber T_{\rm grav} & = & 
 %
 Ic_{\beta}^2c_{\phi}s_{\phi}\partial_{x}^2 V_{\rm grav}-m L h s_{\beta}c_{\beta}s_{\phi}\partial_{x}^{2}V_{\rm grav}
%
 -Ic_{\beta}(c_{\phi}^{2}-s_{\phi}^2)\partial_{x}\partial_{y}V_{\rm grav}\\
 \nonumber &&+m L h s_{\beta}c_{\phi}\partial_{x}\partial_{y}V_{\rm grav}
%
 +2Ic_{\beta}s_{\beta}c_{\phi}s_{\phi}\partial_{x}\partial_{z}V_{\rm grav}+mL h (c_{\beta}^{2}\\
 \nonumber &&-s_{\beta}^{2})s_{\phi}\partial_{x}\partial_{z}V_{\rm grav}
%
  -Ic_{\phi}s_{\phi}\partial_{y}^{2}V_{\rm grav}
 %
 -Is_{\beta}(c_{\phi}^{2}-s_{\phi}^{2})\partial_{y}\partial_{z}V_{\rm grav}\\
 && -mL c_{\beta}c_{\phi}\partial_{y}\partial_{z}V_{\rm grav}
%
 +Is_{\beta}^{2}c_{\phi}s_{\phi}\partial_{z}^{2}V_{\rm grav}+mL h s_{\beta}c_{\beta}s_{\phi}\partial_{z}^{2}V_{\rm grav}\text{.}
\end{eqnarray}
When $\beta$ and $s_{\beta}$ are small, so that $c_{\beta}\approx 1$, the gravitational terms simplify to
\begin{eqnarray}
\nonumber T_{\rm grav} & \approx & \biggl[-I(\partial_{y}^{2}V_{\rm grav}-\partial_{x}^{2}V_{\rm grav})\frac{s_{2\phi}}{2} -I\partial_{x}\partial_{y}V_{\rm grav}c_{2\phi}\\
\nonumber &&+mL h \partial_{x}\partial_{z}V_{\rm grav}s_{\phi}-mL h \partial_{y}\partial_{z}V_{\rm grav}c_{\phi}\biggr]\\
\nonumber && -I\partial_{y}\partial_{z}V_{\rm grav}s_{\beta}c_{2\phi} +mL h\partial_{x}\partial_{y}V_{\rm grav}s_{\beta}c_{\phi}\\
\nonumber &&-m L h \partial_{x}^{2}V_{\rm grav}s_{\beta}s_{\phi}
 +I\partial_{x}\partial_{z}V_{\rm grav}s_{\beta}s_{2\phi} \\
 &&  +mL h\partial_{z}^{2}V_{\rm grav} s_{\beta}s_{\phi}.
\end{eqnarray}
The first five terms (in square brackets) reproduce those of EPF \cite{EPF} (up to a sign), while the remaining five terms do not appear in \cite{EPF}, since they contain $s_{\beta}$ multiplied by gravity gradients which EPF neglected. 
	
	The total torque can then be written as
\begin{eqnarray}
\label{e:T_tot}
\nonumber T_{\rm total} & = &
 \biggr[-I(\partial_{y}^{2}V_{\rm grav}-\partial_{x}^{2}V_{\rm grav})\frac{s_{2\phi}}{2}
 -I\partial_{x}\partial_{y}V_{\rm grav}c_{2\phi}\\
 \nonumber &&+mL h \partial_{x}\partial_{z}V_{\rm grav}s_{\phi}-mL h \partial_{y}\partial_{z}V_{\rm grav}c_{\phi}\biggl]_{\text{EPF}}\\
  \nonumber && +\biggr[q_{+}\frac{I}{4}\left(d_{5yy}-d_{5xx}\right)s_{2\phi}+q_{+}\frac{I}{4}\left(d_{5yx}+d_{5xy}\right)c_{2\phi}\\  
  \nonumber &&-q_{1}mL h d_{5xz}s_{\phi}
 +q_{1}mL h d_{5yz}c_{\phi}- q_{-}mL f_{5z}s_{\beta}s_{\phi}\\
 \nonumber && +q_{+}\frac{I}{4}\left(d_{5yz}+d_{5zy}\right)s_{\beta}c_{2\phi} -q_{1}mL h d_{5yx}s_{\beta}c_{\phi}\\
  \nonumber && +q_{1}mL hd_{5xx} s_{\beta}s_{\phi} -q_{+}\frac{I}{4}\left(d_{5xz}+d_{5zx}\right)s_{\beta}s_{2\phi}\\
\nonumber &&  -q_{1}mL h s_{\beta}s_{\phi}d_{5zz}\biggr]_{\text{EPF+}}\\
\nonumber && +\biggr[-q_{-}mL f_{5x}s_{\phi}
 + q_{-}mL f_{5y}c_{\phi}\biggr]_{\text{EPFx}}\\
&& +\biggl[q_{+}\frac{I}{4}\left(d_{5yx}-d_{5xy}\right) +q_{+}\frac{I}{4}\left(d_{5yz}-d_{5zy}\right)s_{\beta}\biggl]_{\text{NZC}}\text{.}
\end{eqnarray}
To more easily compare our results to those of EPF, we adopt a convention where  in the first set of square brackets,  $[\ ]_{\text{EPF}}$,  are those terms that appear in Eq.~(8) of Ref. \cite{EPF} (the gravity gradient terms).  These are   followed by our three new ``non-EPF'' terms: In Eq.~(\ref{e:T_tot}),  the $[\ ]_{\text{EPF+}}$ terms are the 5th-force equivalents to the gravity gradients. The $[\ ]_{\text{EPF+}}$ terms should reduce to the $[\ ]_{\text{EPF}}$ terms if the 5th force is replaced by the gravitational force. The $[\ ]_{\text{EPFx}}$ terms are entirely new  ($x$ and $y$ direction forces), and the  $[\ ]_{\text{NZC}}$ terms are only relevant if the 5th force has a nonzero curl. Such terms can arise from the interactions of the test samples with a medium, and thus represent a possible new mechanism which might account for the EPF results.

We now introduce the scale value $n$, the constant $n_{0}$, the length $L$, and the torsion constant $\tau$ of the wire and write the torque as
	\begin{equation}
	T_{\rm total} = \tau\frac{n_{0}-n}{2\delta}\text{.}
	\end{equation}
The measured quantity is then
\begin{eqnarray}
\label{e:n0_n}
\nonumber n_{0}-n & = &
\biggl[-\frac{\delta}{\tau}I(\partial_{y}^{2}V_{\rm grav}-\partial_{x}^{2}V_{\rm grav})s_{2\phi}
 -\frac{2\delta}{\tau}I\partial_{x}\partial_{y}V_{\rm grav}c_{2\phi}\\
 \nonumber &&+\frac{2\delta}{\tau}mL h \partial_{x}\partial_{z}V_{\rm grav}s_{\phi}-\frac{2\delta}{\tau}mL h \partial_{y}\partial_{z}V_{\rm grav}c_{\phi}\biggr]_{\text{EPF}}\\
  \nonumber && +\biggl[q_{+}\frac{\delta}{\tau}\frac{I}{2}\left(d_{5yy}-d_{5xx}\right)s_{2\phi}+q_{+}\frac{\delta}{\tau}\frac{I}{2}\left(d_{5yx}+d_{5xy}\right)c_{2\phi}\\  
  \nonumber &&-q_{1}\frac{2\delta}{\tau}mL h d_{5xz}s_{\phi}
 +q_{1}\frac{2\delta}{\tau}mL h d_{5yz}c_{\phi}- q_{-}\frac{2\delta}{\tau}mL f_{5z}s_{\beta}s_{\phi}\\
 \nonumber && +q_{+}\frac{\delta}{\tau}\frac{I}{2}\left(d_{5yz}+d_{5zy}\right)s_{\beta}c_{2\phi} -q_{1}\frac{2\delta}{\tau}mL h d_{5yx}s_{\beta}c_{\phi}\\
  \nonumber && +q_{1}\frac{2\delta}{\tau}mL hd_{5xx} s_{\beta}s_{\phi} -q_{+}\frac{\delta}{\tau}\frac{I}{2}\left(d_{5xz}+d_{5zx}\right)s_{\beta}s_{2\phi}\\
\nonumber &&  -q_{1}\frac{2\delta}{\tau}mL h d_{5zz}s_{\beta}s_{\phi}\biggr]_{\text{EPF+}}\\
\nonumber && + \biggl[-q_{-}\frac{2\delta}{\tau}mL f_{5x}s_{\phi}
 + q_{-}\frac{2\delta}{\tau}mL f_{5y}c_{\phi}\biggr]_{\text{EPFx}}\\
&& +\biggl[q_{+}\frac{\delta}{\tau}\frac{I}{2}\left(d_{5yx}-d_{5xy}\right) +q_{+}\frac{\delta}{\tau}\frac{I}{2}\left(d_{5yz}-d_{5zy}\right)s_{\beta}\biggr]_{\text{NZC}}\text{.}
\end{eqnarray}
	
EPF measured the scale value in four orientations: North (N), East (E), South (S), and West (W). When the apparatus is pointed toward the North, the deflection of the balance from North is simply due to the intrinsic twist $\Delta\phi$ in the wire. Thus, 
\begin{subequations}
\begin{eqnarray}
\phi_{N} & = & \Delta\phi\text{,}\\
s_{\phi_{N}} & = & \Delta\phi\text{,}\\
s_{2\phi_{N}} & = & 2\Delta\phi\text{,}\\
c_{\phi_{N}} & = &  1\text{,}\\
c_{2\phi_{N}} & = & 1\text{.}
\end{eqnarray}
\end{subequations}
Then, 
\begin{eqnarray}
\label{e:n0_nN}
\nonumber n_{0}-n_{N} & = &
\biggl[-\frac{2\delta}{\tau}I(\partial_{y}^{2}V_{\rm grav}-\partial_{x}^{2}V_{\rm grav})\Delta\phi
 -\frac{2\delta}{\tau}I\partial_{x}\partial_{y}V_{\rm grav}\\
 \nonumber &&+\frac{2\delta}{\tau}mL h \partial_{x}\partial_{z}V_{\rm grav}\Delta\phi-\frac{2\delta}{\tau}mL h \partial_{y}\partial_{z}V_{\rm grav}\biggr]_{\text{EPF}}\\
  \nonumber && +\biggr[q_{+}\frac{\delta}{\tau}I\left(d_{5yy}-d_{5xx}\right)\Delta\phi+q_{+}\frac{\delta}{\tau}Id_{5yx}\\  
  \nonumber &&-q_{1}\frac{2\delta}{\tau}mL h d_{5xz}\Delta\phi
 +q_{1}\frac{2\delta}{\tau}mL h d_{5yz}\\
 \nonumber && +q_{+}\frac{\delta}{\tau}Id_{5yz}s_{\beta} -q_{1}\frac{2\delta}{\tau}mL h d_{5yx}s_{\beta}\biggr]_{\text{EPF+}}\\
&& +\biggl[-q_{-}\frac{2\delta}{\tau}mL f_{5x}\Delta\phi
 + q_{-}\frac{2\delta}{\tau}mL f_{5y}\biggr]_{\text{EPFx}}\text{.}
\end{eqnarray}

When the apparatus is pointed in the Eastern direction, the deflection of the beam balance from North is due to a combination of the intrinsic twist, the difference in the measurements $n_{N}$ and $n_{E}$, and a factor of $\pi/2$ due to physically rotating the balance. Thus

\begin{subequations}
\begin{eqnarray}
\phi_{E} & = & \Delta\phi + \frac{n_{N}-n_{E}}{2\delta} + \frac{\pi}{2}\text{,}\\
s_{\phi_{E}} & = & 1\text{,}\\
s_{2\phi_{E}} & = & -2\left(\Delta\phi + \frac{n_{N}-n_{E}}{2\delta}\right)\text{,}\\
c_{\phi_{E}} & = & -\left(\Delta\phi + \frac{n_{N}-n_{E}}{2\delta}\right)\text{,}\\
c_{2\phi_{E}} & = & -1\text{,}
\end{eqnarray}
\end{subequations}
\begin{eqnarray}
\label{e:n0_nE}
\nonumber n_{0}-n_{E} & = &
 \biggl[\frac{2\delta}{\tau}I(\partial_{y}^{2}V_{\rm grav}-\partial_{x}^{2}V_{\rm grav})\left(\Delta\phi + \frac{n_{N}-n_{E}}{2\delta}\right)
 +\frac{2\delta}{\tau}I\partial_{x}\partial_{y}V_{\rm grav}\\
 \nonumber &&+\frac{2\delta}{\tau}mL h \partial_{x}\partial_{z}V_{\rm grav}+\frac{2\delta}{\tau}mL h \partial_{y}\partial_{z}V_{\rm grav}\left(\Delta\phi + \frac{n_{N}-n_{E}}{2\delta}\right)\biggr]_{\text{EPF}}\\
  \nonumber && +\biggl[-q_{+}\frac{\delta}{\tau}I\left(d_{5yy}-d_{5xx}\right)\left(\Delta\phi + \frac{n_{N}-n_{E}}{2\delta}\right)-q_{+}\frac{\delta}{\tau}Id_{5xy}\\  
  \nonumber &&-q_{1}\frac{2\delta}{\tau}mL h d_{5xz}
 -q_{1}\frac{2\delta}{\tau}mL h d_{5yz}\left(\Delta\phi + \frac{n_{N}-n_{E}}{2\delta}\right)\\
 \nonumber && - q_{-}\frac{2\delta}{\tau}mL f_{5z}s_{\beta} -q_{+}\frac{\delta}{\tau}Id_{5zy}s_{\beta}\\
 \nonumber &&+q_{1}\frac{2\delta}{\tau}mL hd_{5xx} s_{\beta}-q_{1}\frac{2\delta}{\tau}mL h d_{5zz}s_{\beta}\biggr]_{\text{EPF+}}\\
&&  +\biggl[-q_{-}\frac{2\delta}{\tau}mL f_{5x}
 - q_{-}\frac{2\delta}{\tau}mL f_{5y}\left(\Delta\phi + \frac{n_{N}-n_{E}}{2\delta}\right)\biggr]_{\text{EPFx}}\text{.}
\end{eqnarray}

Rotating the apparatus to point in the Southern direction yields
\begin{subequations}
\begin{eqnarray}
\phi_{S} & = & \Delta\phi + \frac{n_{N}-n_{S}}{2\delta} + \pi\text{,}\\
s_{\phi_{S}} & = & -\left(\Delta\phi + \frac{n_{N}-n_{S}}{2\delta}\right)\text{,}\\
s_{2\phi_{S}} & = & 2\left(\Delta\phi + \frac{n_{N}-n_{S}}{2\delta}\right)\text{,}\\
c_{\phi_{S}} & = & -1\text{,}\\
c_{2\phi_{S}} & = & 1\text{,}
\end{eqnarray}
\end{subequations}
and, 
\begin{eqnarray}
\label{e:n0_nS}
\nonumber n_{0}-n_{S} & = &
\biggl[-\frac{2\delta}{\tau}I(\partial_{y}^{2}V_{\rm grav}-\partial_{x}^{2}V_{\rm grav})\left(\Delta\phi + \frac{n_{N}-n_{S}}{2\delta}\right)
 -\frac{2\delta}{\tau}I\partial_{x}\partial_{y}V_{\rm grav}\\
 \nonumber &&-\frac{2\delta}{\tau}mL h \partial_{x}\partial_{z}V_{\rm grav}\left(\Delta\phi + \frac{n_{N}-n_{S}}{2\delta}\right)+\frac{2\delta}{\tau}mL h \partial_{y}\partial_{z}V_{\rm grav}\biggr]_{\text{EPF}}\\
  \nonumber && \biggl[+q_{+}\frac{\delta}{\tau}I\left(d_{5yy}-d_{5xx}\right)\left(\Delta\phi + \frac{n_{N}-n_{S}}{2\delta}\right)+q_{+}\frac{\delta}{\tau}Id_{5yx}\\  
  \nonumber &&+q_{1}\frac{2\delta}{\tau}mL h d_{5xz}\left(\Delta\phi + \frac{n_{N}-n_{S}}{2\delta}\right)
 -q_{1}\frac{2\delta}{\tau}mL h d_{5yz}\\
 \nonumber && +q_{+}\frac{\delta}{\tau}Id_{5yz}s_{\beta} +q_{1}\frac{2\delta}{\tau}mL h d_{5yx}s_{\beta}\biggr]_{\text{EPF+}}\\
 && +\biggl[q_{-}\frac{2\delta}{\tau}mL f_{5x}\left(\Delta\phi + \frac{n_{N}-n_{S}}{2\delta}\right)
 - q_{-}\frac{2\delta}{\tau}mL f_{5y}\biggr]_{\text{EPFx}}\text{.}
\end{eqnarray}

Finally, rotating the apparatus to point in the Western direction yields
\begin{subequations}
\begin{eqnarray}
\phi_{W} & = & \Delta\phi + \frac{n_{N}-n_{W}}{2\delta} + \frac{3\pi}{2}\text{,}\\
s_{\phi_{W}} & = & -1\text{,}\\
s_{2\phi_{W}} & = & -2\left(\Delta\phi + \frac{n_{N}-n_{W}}{2\delta}\right)\text{,}\\
c_{\phi_{W}} & = & \left(\Delta\phi + \frac{n_{N}-n_{W}}{2\delta}\right)\text{,}\\
c_{2\phi_{W}} & = & -1\text{,}
\end{eqnarray}
\end{subequations}
and,
\begin{eqnarray}
\label{e:n0_nW}
\nonumber n_{0}-n_{W} & = &
 \biggl[\frac{2\delta}{\tau}I(\partial_{y}^{2}V_{\rm grav}-\partial_{x}^{2}V_{\rm grav})\left(\Delta\phi + \frac{n_{N}-n_{W}}{2\delta}\right)
 +\frac{2\delta}{\tau}I\partial_{x}\partial_{y}V_{\rm grav}\\
 \nonumber &&-\frac{2\delta}{\tau}mL h \partial_{x}\partial_{z}V_{\rm grav}-\frac{2\delta}{\tau}mL h \partial_{y}\partial_{z}V_{\rm grav}\left(\Delta\phi + \frac{n_{N}-n_{W}}{2\delta}\right)\biggr]_{\text{EPF}}\\
  \nonumber && +\biggl[-q_{+}\frac{\delta}{\tau}I\left(d_{5yy}-d_{5xx}\right)\left(\Delta\phi + \frac{n_{N}-n_{W}}{2\delta}\right)-q_{+}\frac{\delta}{\tau}Id_{5xy}\\
  \nonumber && +q_{1}\frac{2\delta}{\tau}mL h d_{5xz}+q_{1}\frac{2\delta}{\tau}mL h d_{5yz}\left(\Delta\phi + \frac{n_{N}-n_{W}}{2\delta}\right)\\
  \nonumber && + q_{-}\frac{2\delta}{\tau}mL f_{5z}s_{\beta}-q_{+}\frac{\delta}{\tau}Id_{5zy}s_{\beta}\\  
  \nonumber &&-q_{1}\frac{2\delta}{\tau}mL hd_{5xx} s_{\beta}+q_{1}\frac{2\delta}{\tau}mL h d_{5zz}s_{\beta}\biggl]_{\text{EPF+}}\\
&&  +\biggl[q_{-}\frac{2\delta}{\tau}mL f_{5x}
 + q_{-}\frac{2\delta}{\tau}mL f_{5y}\left(\Delta\phi + \frac{n_{N}-n_{W}}{2\delta}\right)\biggr]_{\text{EPFx}}\text{.}
\end{eqnarray}
	
To eliminate the contributions from gravity gradients, we combine measurements by introducing $\sigma\equiv n_{N}-n_{S}$, where	
\begin{eqnarray}
\label{e:sigma}
\nonumber \sigma & = &
 \biggl[-\frac{2\delta}{\tau}I(\partial_{y}^{2}V_{\rm grav}-\partial_{x}^{2}V_{\rm grav})\left(\frac{n_{N}-n_{S}}{2\delta}\right)\\
 \nonumber &&-\frac{2\delta}{\tau}mL h \partial_{x}\partial_{z}V_{\rm grav}\left(2\Delta\phi + \frac{n_{N}-n_{S}}{2\delta}\right)+\frac{4\delta}{\tau}mL h \partial_{y}\partial_{z}V_{\rm grav}\biggr]_{\text{EPF}}\\
  \nonumber && +\biggl[q_{+}\frac{\delta}{\tau}I\left(d_{5yy}-d_{5xx}\right)\left(\frac{n_{N}-n_{S}}{2\delta}\right)+q_{1}\frac{4\delta}{\tau}mL h d_{5yx}s_{\beta}\\
  \nonumber &&+q_{1}\frac{2\delta}{\tau}mL h d_{5xz}\left(2\Delta\phi + \frac{n_{N}-n_{S}}{2\delta}\right)
 -q_{1}\frac{4\delta}{\tau}mL h d_{5yz}\biggr]_{\text{EPF+}}\\
\nonumber && +\biggl[q_{-}\frac{2\delta}{\tau}mL f_{5x}\left(2\Delta\phi + \frac{n_{N}-n_{S}}{2\delta}\right)
 - q_{-}\frac{4\delta}{\tau}mL f_{5y}\biggr]_{\text{EPFx}}\\
\nonumber & = & \frac{4\delta}{\tau}mL h \left(\partial_{y}\partial_{z}V_{\rm grav}-q_{1}d_{5yz}-q_{-}\frac{f_{5y}}{h}\right)\\
\nonumber &&-\frac{2\delta}{\tau}I\left[(\partial_{y}^{2}V_{\rm grav}-\partial_{x}^{2}V_{\rm grav})-\frac{q_{+}}{2}(d_{5yy}-d_{5xx})\right]\frac{n_{N}-n_{S}}{2\delta}\\
\nonumber &&-\frac{2\delta}{\tau}mL h \left(\partial_{x}\partial_{z}V_{\rm grav}-q_{1}d_{5xz}-q_{-}\frac{f_{5x}}{h}\right)\left(2\Delta\phi+\frac{n_{N}-n_{S}}{2\delta}\right)\\
 &&+q_{1}\frac{4\delta}{\tau}mL h s_{\beta}d_{5yx}\text{.}
\end{eqnarray}
Similarly, for $\nu \equiv n_{E}-n_{W}$, 
\begin{eqnarray}
\label{e:nu}
\nonumber \nu & = &
 \biggl[\frac{2\delta}{\tau}I(\partial_{y}^{2}V_{\rm grav}-\partial_{x}^{2}V_{\rm grav})\left(\frac{n_{E}-n_{W}}{2\delta}\right)\\
 \nonumber &&-\frac{4\delta}{\tau}mL h \partial_{x}\partial_{z}V_{\rm grav}-\frac{2\delta}{\tau}mL h \partial_{y}\partial_{z}V_{\rm grav}\left(2\Delta\phi + \frac{2n_{N}-n_{E}-n_{W}}{2\delta}\right)\biggr]_{\text{EPF}}\\
  \nonumber && +\biggl[-q_{+}\frac{\delta}{\tau}I\left(d_{5yy}-d_{5xx}\right)\left(\frac{n_{E}-n_{W}}{2\delta}\right)+ q_{-}\frac{4\delta}{\tau}mL f_{5z}s_{\beta}\\  
  \nonumber &&+q_{1}\frac{4\delta}{\tau}mL h d_{5xz}
 +q_{1}\frac{2\delta}{\tau}mL h d_{5yz}\left(2\Delta\phi + \frac{2n_{N}-n_{E}-n_{W}}{2\delta}\right)\\
 \nonumber && -q_{1}\frac{4\delta}{\tau}mL hd_{5xx} s_{\beta}+q_{1}\frac{4\delta}{\tau}mL h d_{5zz}s_{\beta}\biggr]_{\text{EPF+}}\\
\nonumber &&  +\biggl[q_{-}\frac{4\delta}{\tau}mL f_{5x}
 + q_{-}\frac{2\delta}{\tau}mL f_{5y}\left(2\Delta\phi + \frac{2n_{N}-n_{E}-n_{W}}{2\delta}\right)\biggr]_{\text{EPFx}}\\
\nonumber & = & -\frac{4\delta}{\tau}mL h \left(\partial_{x}\partial_{z}V_{\rm grav}-q_{1}d_{5xz}-q_{-}\frac{f_{5x}}{h}\right)\\
\nonumber &&+\frac{2\delta}{\tau}I\left[(\partial_{y}^{2}V_{\rm grav}-\partial_{x}^{2}V_{\rm grav})-\frac{q_{+}}{2}(d_{5yy}-d_{5xx})\right]\frac{n_{E}-n_{W}}{2\delta}\\
\nonumber &&-\frac{2\delta}{\tau}mL h \left(\partial_{y}\partial_{z}V_{\rm grav}-q_{1}d_{5yz}-q_{-}\frac{f_{5y}}{h}\right)\left(2\Delta\phi+\frac{2n_{N}-n_{E}-n_{W}}{2\delta}\right)\\
&& + q_{-}\frac{4\delta}{\tau}mL s_{\beta}f_{5z}+q_{1}\frac{4\delta}{\tau}mL h s_{\beta}d_{5zz}-q_{1}\frac{4\delta}{\tau}mL h s_{\beta}d_{5xx}\text{.}
	\end{eqnarray}	
	
Equations~(\ref{e:sigma}) and (\ref{e:nu}) reproduce Eqs. (14) and (15) in the original EPF work\cite{EPF}. In the final step of both equations, we have deviated from our convention of grouping the E\"otv\"os contributions at the beginning of the equation, to grouping the terms in such a way as to simplify the next step of the analysis, the computation of $\nu/\sigma$. In Ref \cite{EPF}, $\nu$ contained the WEP-violating term of interest along with gravity-gradient terms, while $\sigma$ contained only gravity-gradients. To remove the gravity gradient contributions, EPF divided $\nu$ by $\sigma$. Eqs.~(\ref{e:sigma}) and (\ref{e:nu}) can be further simplified by defining
\begin{eqnarray}
\label{e:sigma_par}
\tau\frac{\sigma}{2\delta} & = & C_{1}-C_{2}\frac{n_{N}-n_{S}}{2\delta}-C_{3}\left(2\Delta\phi+\frac{n_{N}-n_{S}}{2\delta}\right)+C_{4}s_{\beta}\text{,}\\
\label{e:nu_par}
\tau\frac{\nu}{2\delta} & = & -D_{1}+D_{2}\frac{n_{E}-n_{W}}{2\delta}-D_{3}\left(2\Delta\phi+\frac{2n_{N}-n_{E}-n_{W}}{2\delta}\right)+ D_{4}s_{\beta}\text{,}
\end{eqnarray}	
with
\begin{eqnarray}
C_{1} & = & 2mL h \left(\partial_{y}\partial_{z}V_{\rm grav}-q_{1}d_{5yz}-q_{-}\frac{f_{5y}}{h}\right)\text{,}\\
C_{2} & = & I\left[(\partial_{y}^{2}V_{\rm grav}-\partial_{x}^{2}V_{\rm grav})-\frac{q_{+}}{2}(d_{5yy}-d_{5xx})\right]\text{,}\\
C_{3} & = & mL h \left(\partial_{x}\partial_{z}V_{\rm grav}-q_{1}d_{5xz}-q_{-}\frac{f_{5x}}{h}\right)\text{,}\\
C_{4} & = & 2q_{1}mL h d_{5yx}\text{,}\\
\nonumber D_{1} & = & 2mL h \left(\partial_{x}\partial_{z}V_{\rm grav}-q_{1}d_{5xz}-q_{-}\frac{f_{5x}}{h}\right)\\
& = & 2C_{3}\text{,}\\
\nonumber D_{2} & = & I\left[(\partial_{y}^{2}V_{\rm grav}-\partial_{x}^{2}V_{\rm grav})-\frac{q_{+}}{2}(d_{5yy}-d_{5xx})\right]\\
 & = & C_{2}\text{,}\\
\nonumber D_{3} & = & mL h \left(\partial_{y}\partial_{z}V_{\rm grav}-q_{1}d_{5yz}-q_{-}\frac{f_{5y}}{h}\right)\\
 & = & \frac{1}{2}C_{1}\text{,}\\
D_{4} & = & 2q_{-}mL f_{5z}+2q_{1}mL h d_{5zz}-2q_{1}mL h d_{5xx}\text{.}
\end{eqnarray}
The definitions of the $C$'s and $D$'s have been chosen so that the $C_{i}$ and $D_{i}$ would all have the same approximate magnitude. We assume, for the moment, that the $d_{5}$'s are roughly the same order of magnitude as the gravity gradients, and that $\vec{f}_{5}/h$ is also of the same magnitude.  Since $h$ and $L$ are of the same order, $I$ and $mLh$ will be as well. Returning to Eqs.~(\ref{e:sigma_par}) and (\ref{e:nu_par}), we know that $\Delta \alpha$ and $s_{\beta}$ are small, and that $\delta$ is larger than $n_{N}$, $n_{E}$, $n_{S}$, and $n_{W}$. Hence the lowest order approximations of $\sigma$ and $\nu$ are given by,
\begin{eqnarray}
\label{e:sigma_NLO}
\tau\frac{\sigma}{2\delta} &\approx& C_{1} = 2mL h \left(\partial_{y}\partial_{z}V_{\rm grav}-q_{1}d_{5yz}-q_{-}\frac{f_{5y}}{h}\right)\text{,}\\
\label{e:nu_NLO}
\tau\frac{\nu}{2\delta} &\approx& -D_{1} = -2mL h \left(\partial_{x}\partial_{z}V_{\rm grav}-q_{1}d_{5xz}-q_{-}\frac{f_{5x}}{h}\right)\text{.}
\end{eqnarray}

It follows from the preceding discussion that our analog of the EPF expression for $\nu/\sigma$ is given by
\begin{eqnarray}
\label{e:nu_sig1}
\nonumber \frac{\nu}{\sigma} & = & \frac{-D_{1}+C_{2}\frac{n_{E}-n_{W}}{2\delta}-C_{1}\left(\Delta\phi+\frac{2n_{N}-n_{E}-n_{W}}{4\delta}\right)+ D_{4}s_{\beta}}{C_{1}-C_{2}\frac{n_{N}-n_{S}}{2\delta}-D_{1}\left(\Delta\phi+\frac{n_{N}-n_{S}}{4\delta}\right)+C_{4}s_{\beta}}\\
\nonumber && \\
\nonumber  & = & \frac{-D_{1}+C_{2}\frac{n_{E}-n_{W}}{2\delta}-C_{1}\left(\Delta\phi+\frac{2n_{N}-n_{E}-n_{W}}{1\delta}\right)+ D_{4}s_{\beta}}{C_{1}\left(1-\frac{C_{2}}{C_{1}}\frac{n_{N}-n_{S}}{2\delta}-\frac{D_{1}}{C_{1}}\left(\Delta\phi+\frac{n_{N}-n_{S}}{4\delta}\right)+\frac{C_{4}}{C_{1}}s_{\beta}\right)}\\
\nonumber && \\
\nonumber & \approx & \left(-\frac{D_{1}}{C_{1}}+\frac{C_{2}}{C_{1}}\frac{n_{E}-n_{W}}{2\delta}-\left(\Delta\phi+\frac{2n_{N}-n_{E}-n_{W}}{4\delta}\right)+ \frac{D_{4}}{C_{1}}s_{\beta}\right)\\
\nonumber && \\
\nonumber &&\times\left(1+\frac{C_{2}}{C_{1}}\frac{n_{N}-n_{S}}{2\delta} +\frac{D_{1}}{C_{1}}\left(\Delta\phi+\frac{n_{N}-n_{S}}{4\delta}\right)-\frac{C_{4}}{C_{1}}s_{\beta}\right)\\
\nonumber && \\
\nonumber & \approx & -\frac{D_{1}}{C_{1}}+\frac{C_{2}}{C_{1}}\frac{n_{E}-n_{W}}{2\delta}-\left(\Delta\phi+\frac{2n_{N}-n_{E}-n_{W}}{4\delta}\right)+ \frac{D_{4}}{C_{1}}s_{\beta}\\
\nonumber && \\
  &&-\frac{D_{1}}{C_{1}}\frac{C_{2}}{C_{1}}\frac{n_{N}-n_{S}}{2\delta} -\left(\frac{D_{1}}{C_{1}}\right)^{2}\left(\Delta\phi+\frac{n_{N}-n_{S}}{4\delta}\right)+\frac{D_{1}}{C_{1}}\frac{C_{4}}{C_{1}}s_{\beta}\text{.}
\end{eqnarray}

We can now use our approximations (\ref{e:sigma_NLO}) and (\ref{e:nu_NLO}) to simplify Eq.~(\ref{e:nu_sig1}). In every term except the first, $D_{1}/C_{1}$ can be replaced by $-\nu/\sigma$, and $C_{1}$ by $\tau\sigma/2\delta$. (These substitutions serve to express $C_{1}$ and  $D_{1}/C_{1}$ in terms of the measured quantities $\sigma$ and $\nu$). Using $\sigma = n_{N}-n_{S}$ and $\nu = n_{E}-n_{W}$, we can now write
\begin{eqnarray}
	\label{e:nu_sigma_1}
\nonumber \frac{\nu}{\sigma} 
& = & -\frac{D_{1}}{C_{1}}+\frac{2C_{2}}{\tau}\frac{\nu}{\sigma}-\left(\frac{\nu^{2}}{\sigma^{2}}+1\right)\Delta\phi-\frac{2n_{N}-n_{E}-n_{W}}{4\delta}\\
&&-\frac{\nu^{2}}{\sigma^{2}}\frac{\sigma}{4\delta}+ \frac{2\delta}{\sigma\tau}D_{4}s_{\beta}-\frac{\nu}{\sigma}\frac{2\delta}{\sigma\tau}C_{4}s_{\beta}\text{.}
\end{eqnarray}
Eq.~(\ref{e:nu_sigma_1}) can be further simplified by noting that since $f_{5i}/h$ and $d_{5ij}$ are presumed to be much smaller than the gravity gradient terms, we can approximate $\sigma$ and $\nu$ as simply the gravity gradient terms to lowest order, 
\begin{eqnarray}
\label{e:sigma_lowest}
\tau\frac{\sigma}{2\delta} &\approx& 2mL h\partial_{y}\partial_{z}V_{\rm grav}\text{,}\\
\label{e:nu_lowest}
\tau\frac{\nu}{2\delta} &\approx& -2mL h \partial_{x}\partial_{z}V_{\rm grav}\text{.}
\end{eqnarray}

Similarly, the first term in Eq.~(\ref{e:nu_sigma_1}) can be approximated as,
\begin{eqnarray}
\nonumber \frac{D_{1}}{C_{1}} & = & \frac{\left(\partial_{x}\partial_{z}V_{\rm grav}-q_{1}d_{5xz}-q_{-}\frac{f_{5x}}{h}\right)}{ \left(\partial_{y}\partial_{z}V_{\rm grav}-q_{1}d_{5yz}-q_{-}\frac{f_{5y}}{h}\right)}\\
\nonumber & =  &\frac{\left(\partial_{x}\partial_{z}V_{\rm grav}-q_{1}d_{5xz}-q_{-}\frac{f_{5x}}{h}\right)}{ \partial_{y}\partial_{z}V_{\rm grav}}\left(1-\frac{q_{1}d_{5yz}+q_{-}\frac{f_{5y}}{h}}{\partial_{y}\partial_{z}V_{\rm grav}}\right)^{-1}\\
\nonumber&\approx &\frac{\partial_{x}\partial_{z}V_{\rm grav}}{\partial_{y}\partial_{z}V_{\rm grav}}-\frac{q_{1}d_{5xz}+q_{-}\frac{f_{5x}}{h}}{\partial_{y}\partial_{z}V_{\rm grav}} +\frac{\partial_{x}\partial_{z}V_{\rm grav}}{\partial_{y}\partial_{z}V_{\rm grav}}\frac{q_{1}d_{5yz}+q_{-}\frac{f_{5y}}{h}}{\partial_{y}\partial_{z}V_{\rm grav}}\\
\nonumber &\approx & \frac{\partial_{x}\partial_{z}V_{\rm grav}}{\partial_{y}\partial_{z}V_{\rm grav}}-q_{1}\frac{1}{\partial_{y}\partial_{z}V_{\rm grav}}\left(d_{5xz}-\frac{\partial_{x}\partial_{z}V_{\rm grav}}{\partial_{y}\partial_{z}V_{\rm grav}}d_{5yz}\right)\\
 &&+\frac{q_{-}}{h}\frac{1}{\partial_{y}\partial_{z}V_{\rm grav}}\left(f_{5x}-\frac{\partial_{x}\partial_{z}V_{\rm grav}}{\partial_{y}\partial_{z}V_{\rm grav}}f_{5y}\right)\text{.}
	\end{eqnarray}
Using Eq.~(\ref{e:sigma_lowest}), $\left(\partial_{y}\partial_{z}V_{\rm grav}\right)^{-1}$ can be expressed in terms of $\sigma$, and $\left(\partial_{x}\partial_{z}V_{\rm grav}/\partial_{y}\partial_{z}V_{\rm grav}\right)$ can be replaced by $\left(-\nu/\sigma\right)$ in the second and third terms. It then follows that
\begin{eqnarray}
\label{e:NLO_exp}
\nonumber \frac{D_{1}}{C_{1}}
 & \approx & \frac{\partial_{x}\partial_{z}V_{\rm grav}}{\partial_{y}\partial_{z}V_{\rm grav}}-q_{1}\frac{4\delta}{\sigma\tau}mLh\left(d_{5xz}+\frac{\nu}{\sigma}d_{5yz}\right)\\
 &&+q_{-}\frac{4\delta}{\sigma\tau}mL\left(f_{5x}+\frac{\nu}{\sigma}f_{5y}\right)\text{.}
	\end{eqnarray}

Combining Eqs.~(\ref{e:nu_sigma_1}) and (\ref{e:NLO_exp}) we then find,
	\begin{eqnarray}
\nonumber \frac{\nu}{\sigma} 
& = &\biggl[-\frac{\partial_{x}\partial_{z}V_{\rm grav}}{\partial_{y}\partial_{z}V_{\rm grav}}+\frac{2I}{\tau}\frac{\nu}{\sigma}\left(\partial_{y}^{2}V_{\rm grav}-\partial_{x}^{2}V_{\rm grav}\right)-\left(\frac{\nu^{2}}{\sigma^{2}}+1\right)\Delta\phi\\
\nonumber &&-\frac{\nu^{2}}{\sigma^{2}}\frac{\sigma}{4\delta}-\frac{2n_{N}-n_{E}-n_{W}}{4\delta}\biggr]_{\text{EPF}}\\
\nonumber   &&  +\biggl[ q_{-}\frac{4\delta}{\sigma\tau}mL s_{\beta}f_{5z}+q_{1}\frac{4\delta}{\sigma\tau}mL h\left(d_{5xz}+\frac{\nu}{\sigma}d_{5yz}\right)\\
\nonumber && -q_{+}\frac{I}{\tau}\frac{\nu}{\sigma}\left(d_{5yy}-d_{5xx}\right)-q_{1}\frac{4\delta}{\sigma\tau}mL h s_{\beta}\left(d_{5xx}+\frac{\nu}{\sigma}d_{5yx}-d_{5zz}\right)\biggr]_{\text{EPF+}}\\
&&+ \biggl[ q_{-}\frac{4\delta}{\sigma\tau}mL \left(f_{5x}+\frac{\nu}{\sigma}f_{5y}\right)\biggr]_{\text{EPFx}}.
\label{nu/sigma}
	\end{eqnarray}	
In Eq.~(\ref{nu/sigma})  we have returned to our convention of grouping the EPF terms first. To eliminate the surviving gradient contributions, the sample $q_{1}$ is replaced with the sample $q_{1}^{\prime}$ and the measurements are repeated, leading to a new term $\nu^{\prime} / \sigma^{\prime}$. We can then cancel the gravity gradient contributions to lowest working order by noting that 

\begin{eqnarray}
\label{e:nusigdif_new}
\nonumber\frac{\nu}{\sigma} - \frac{\nu^{\prime}}{\sigma^{\prime}}
 & = & \biggl[-\left(\frac{\nu^{2}}{\sigma^{2}}+1\right)(\Delta\phi-\Delta\phi^{\prime})\biggr]\\
\nonumber &&+ \biggl[\Delta q\frac{4\delta}{\sigma\tau}mL s_{\beta}f_{5z}\biggr]\\
\nonumber && +\biggl[\Delta q\frac{4\delta}{\sigma\tau}mL \left(f_{5x}+\frac{\nu}{\sigma}f_{5y}\right)\\
\nonumber &&-\Delta q\frac{I}{\tau}\frac{\nu}{\sigma}(d_{5yy}-d_{5xx})\\
\nonumber &&+\Delta q\frac{4\delta}{\sigma\tau}mL h\left(d_{5xz}+\frac{\nu}{\sigma}d_{5yz}\right)\\
&& -\Delta q\frac{4\delta}{\sigma\tau}mL h s_{\beta}\left(d_{5xx}+\frac{\nu}{\sigma}d_{5yx}-d_{5zz}\right)\biggr]\text{.}
\end{eqnarray}
In Eq.~(\ref{e:nusigdif_new}), $\Delta q = q_{1}-q_{1}^{\prime}$, and the first bracket corresponds to Eq.~(18) of EPF\cite{EPF}. The objective of the EPF experiment was to measure the differences of the WEP-violating parameters for the different substances, referred to as $\kappa_{a}-\kappa_{a}^{\prime}$, and from Eq.~(19) of \cite{EPF}, 
   \begin{eqnarray} 
\kappa_{a}-\kappa_{a}^{\prime} & = & \frac{\sigma\tau}{4\sigma mLg s_{\beta}}\left[\left(\frac{\nu}{\sigma}-\frac{\nu^{\prime}}{\sigma^{\prime}}\right)+ \left(\frac{\nu^{2}}{\sigma^{2}}+1\right)\left(\Delta\phi-\Delta\phi^{\prime}\right)\right]\text{.}
\end{eqnarray}
This parameter, which is represented by $\eta$ in modern works \cite{Will2018}, and referred to as the ``E\"otv\"os Parameter,'' can be written in terms of our 5th force by combining Eq.~(19) of \cite{EPF} with Eq.~(\ref{e:nusigdif_new}) above:
\begin{eqnarray}
\nonumber \eta_{\rm EPF} & = & \Delta\kappa s_{\beta} = \frac{\sigma\tau}{4\sigma mLg }\left[\left(\frac{\nu}{\sigma}-\frac{\nu^{\prime}}{\sigma^{\prime}}\right)+\left(\frac{\nu^{2}}{\sigma^{2}}+1\right)\left(\Delta\phi-\Delta\phi^{\prime}\right)\right]\\
\nonumber  & = & \Delta q\frac{f_{5z}}{g}s_{\beta} +\Delta q\frac{1}{g } \left(f_{5x}+\frac{\nu}{\sigma}f_{5y}\right)\\
\nonumber &&-\Delta q \frac{L}{g }\frac{\nu}{2\delta}(d_{5yy}-d_{5xx})\\
\nonumber &&+\Delta q\frac{h}{g }\left(d_{5xz}+\frac{\nu}{\sigma}d_{5yz}\right)\\
&& -\Delta q\frac{h}{g}s_{\beta}\left(d_{5xx}+\frac{\nu}{\sigma}d_{5yx}-d_{5zz}\right)\text{.}
\label{eta EPF}
\end{eqnarray}
As in the discussion following Eq.~(\ref{e:rawfifthforce}), $\Delta q$ is the difference in the 5th force charges of the samples, where $q = B/\mu$, $B$ is baryon number, and $\mu$ is mass in units of $m(^{1}$H$_{1})$.
	
\section{Application to the E\"ot-Wash Experiment}
\label{section:Eot-Wash}

\begin{figure}[t]
  \centering
  \includegraphics[width=0.6\columnwidth]{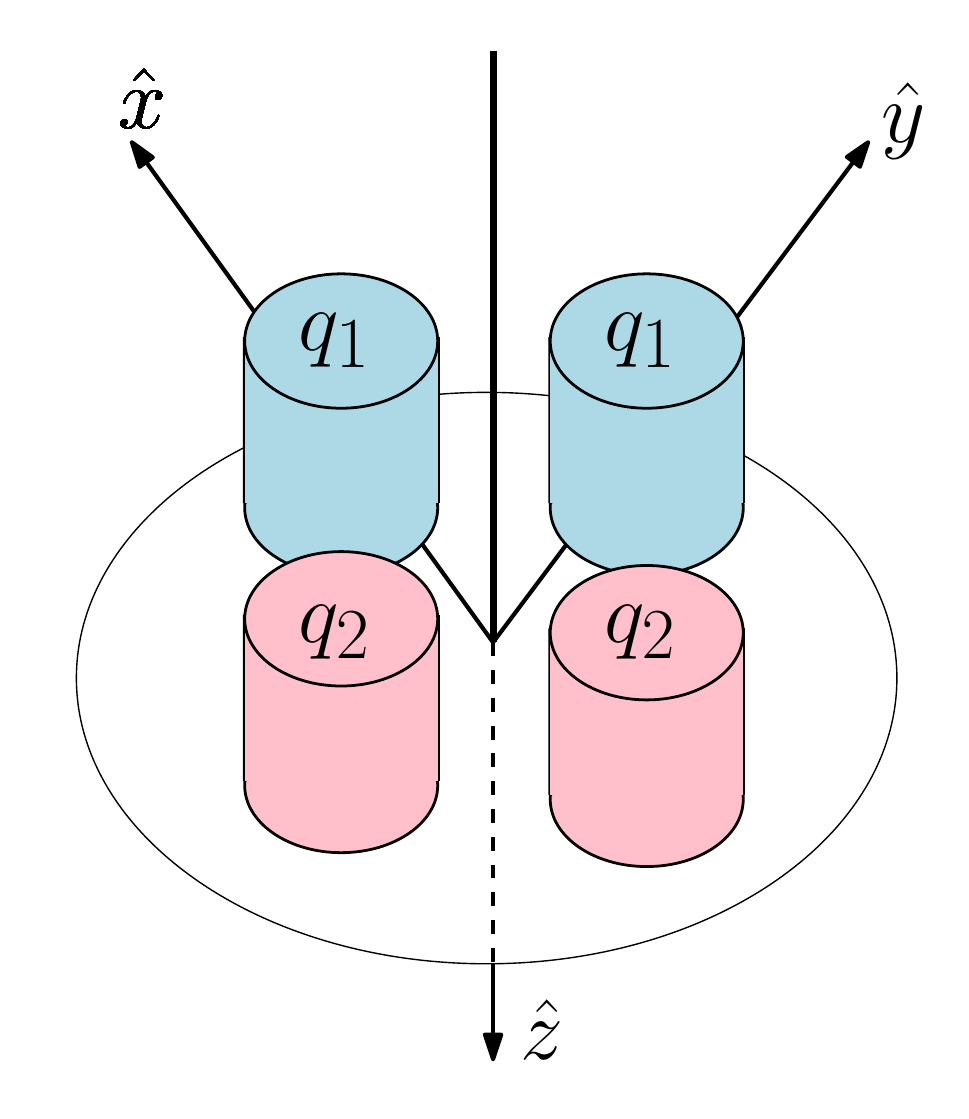}
  \caption{Test mass arrangement of the E\"ot-Wash Experiment \cite{Eotwash1987}.}
  \label{fi:EotWash}
\end{figure}

 For illustrative purposes, we next apply the preceding formalism  to the E\"ot-Wash (E-W) experiment of Stubbs et al. \cite{Eotwash1987}, which employed four masses, two of each sample 1 and 2. They form two perpendicular 5th force dipoles and an overall mass quadrupole, as illustrated in Fig.~\ref{fi:EotWash}. 

The product of the fifth-force charge $q(\vec{r})$ and the mass density $\rho(\vec{r})$ can be written as
\begin{eqnarray}
\nonumber q\left(\vec{r}\right)\rho\left(\vec{r}\right) & = & q_{1}m\left[\delta\left(x-L\right)\delta\left(y\right)\delta\left(z\right) + \delta\left(x\right)\delta\left(y-L\right)\delta\left(z\right)\right]\\
&&+q_{2}m \left[\delta\left(x+L\right)\delta\left(y\right)\delta\left(z\right) + \delta\left(x\right)\delta\left(y+L\right)\delta\left(z\right)\right]\text{.}
\end{eqnarray}
Starting from Eq.~(\ref{e:T_tot}) for the torque, we let $\phi\rightarrow \phi+\pi/2$, and add the result to the original expression. If we then set $h=0$, the 5th force torque, $T_{5}$ is given by, 

\begin{eqnarray}
\nonumber T_{5} & = &
 %
-(q_{1}-q_{2})mL (s_{\phi}+c_{\phi})f_{5x}
 - (q_{1}-q_{2})mL (s_{\phi}-c_{\phi})f_{5y}
 - (q_{1}-q_{2})mL s_{\beta}(s_{\phi}+c_{\phi})f_{5z}\\
\nonumber && -(q_{1}+q_{2})\frac{I}{2}(d_{5yx}-d_{5xy})
 +(q_{1}+q_{2})\frac{I}{2}s_{\beta}(d_{5yz}-d_{5zy})\\
\nonumber & = & -(q_{1}-q_{2})mL \sqrt{2}\left(f_{5x}+s_{\beta}f_{5z}\right)\cos\left({\phi-\frac{\pi}{4}}\right)
 - (q_{1}-q_{2})mL \sqrt{2}f_{5y}\sin\left({\phi-\frac{\pi}{4}}\right)\\
&& -(q_{1}+q_{2})\frac{I}{2}(d_{5yx}-d_{5xy})
 +(q_{1}+q_{2})\frac{I}{2}s_{\beta}(d_{5yz}-d_{5zy})\text{.}
	\end{eqnarray}
To eliminate the contribution from any intrinsic twist in the fiber we define $\overline{T}(\phi) = T_{5}(\phi)-T_{5}(\phi+\pi)$, where
\begin{eqnarray}
\label{e:Tbar}
\nonumber \overline{T} & = & -(q_{1}-q_{2})mL 2\sqrt{2}\left(f_{5x}+s_{\beta}f_{5z}\right)\cos\left({\phi-\frac{\pi}{4}}\right)
 - (q_{1}-q_{2})mL 2\sqrt{2}f_{5y}\sin\left({\phi-\frac{\pi}{4}}\right)\\
 & = & -(q_{1}-q_{2})mL 2\sqrt{2}\sqrt{\left(f_{5x}+s_{\beta}f_{5z}\right)^{2}+f_{5y}^2}\cos\left(\phi+\delta\right)\text{,}\\
 \delta & = & \tan^{-1}\left(\frac{f_{5y}}{f_{5x}+s_{\beta}f_{5z}}\right)-\frac{\pi}{4}\text{.}
 \end{eqnarray}
In Eq.~(\ref{e:Tbar}) the acceleration difference is $\Delta a = (q_{1}-q_{2})\sqrt{2}\sqrt{\left(f_{5x}+s_{\beta}f_{5z}\right)^{2}+f_{5y}^2}$, so  the E\"otv\"os parameter for the E-W experiment is
\begin{equation}
 \eta_{\rm EW} = \frac{\Delta a}{g} = \Delta q\frac{\sqrt{2}}{g}\sqrt{\left(f_{5x}+s_{\beta}f_{5z}\right)^{2}+f_{5y}^2}.
 \label{eta EW}
 \end{equation}
 
\section{Implications for 5th force Models}
\label{sec:Implications}

We can now contrast the signals from the general 5th force given in Sec.~\ref{section:Methodology} as they would appear in the EPF and E-W experiments by comparing the \Eotvos\ parameters,  Eqs.~(\ref{eta EPF}) and (\ref{eta EW}), for these experiments:
\begin{eqnarray}
\eta_{\rm EW} & = & \frac{\Delta a}{g} = \Delta q\frac{\sqrt{2}}{g}\sqrt{\left(f_{5x}+s_{\beta}f_{5z}\right)^{2}+f_{5y}^2},  \label{eta EW 2} 
\\
 \eta_{\rm EPF}  & = & \Delta q\frac{f_{5z}}{g}s_{\beta} +\Delta q\frac{1}{g } \left(f_{5x}+\frac{\nu}{\sigma}f_{5y}\right)  \nonumber \\
&&-\Delta q \frac{L}{g }\frac{\nu}{2\delta}(d_{5yy}-d_{5xx}) \nonumber\\
\nonumber &&+\Delta q\frac{h}{g }\left(d_{5xz}+\frac{\nu}{\sigma}d_{5yz}\right)\\
&& -\Delta q\frac{h}{g}s_{\beta}\left(d_{5xx}+\frac{\nu}{\sigma}d_{5yx}-d_{5zz}\right). \label{eta EPF 2} 
\end{eqnarray}
The differences between $\eta_{\rm EW}$ and $ \eta_{\rm EPF}$  are evident.  While both experiments are sensitive to the coefficients $f_{5i}$, the functional dependence appearing in $\eta_{\rm EW}$ differs considerably from that in $\eta_{\rm EPF}$.  Furthermore, the EPF experiment is sensitive to a 5th force that depends on $d_{5ij}$ (force gradients), while these  terms are absent from $\eta_{\rm EW}$.  This vividly illustrates an important point:  while the EPF and E-W experiments both use torsion balances, their different configurations of test masses and methodologies make them potentially sensitive to different signals.    It is thus clearly possible for a new force to be detected by one experiment and be missed by the other.  In the next section, we will explore these differences in greater detail.

\section{Examples of General Forces and their Effects on the EPF and E-W Experiments}
\label{sec:Examples}

\subsection{Scalar-Vector Model}
\label{sec:SV_Model}

Since it is not the purpose of this paper to introduce new models of a 5th force, in order to  demonstrate the differences in sensitivity to new forces by the EPF and E-W experiments, in this section we will use a simple scalar-vector model for illustrative purposes only.  The preceding results suggest a possible set of requirements for a 5th force model aimed at accounting for the EPF data in light of the null result from the E-W experiment \cite{Eotwash1987}, and later variants \cite{FischbachBook}:
\begin{enumerate}
\item The interaction should be proportional to baryon number $B$.
\item The torque due to the force, $f_{5i}$, should vanish in the E-W experiment.
\item The torque due to the 5th force gradients, $d_{5ij}$, should be nonzero in the EPF experiment.
\item If the new interaction arises from terrestrial sources, it  should be of relatively short range, $\lambda$ on the order of $1 \text{ m} \lesssim \lambda \lesssim 1 \text{ km}$ where constraints from existing experiments are comparatively weak.
\end{enumerate}
The first requirement is straightforward to implement.   The second  and third requirements can be achieved if the  force is small at some points, but not at others, which is the same as requiring that the potential have minima at some points. As an example, consider the quadratic potential $V = \frac{1}{2}m\omega^{2}\left(x-x_{0}\right)^{2}$, which has a minimum at $x=x_{0}$. The force, $F_{x}=m\omega^{2}\left(x-x_{0}\right)$, vanishes at $x_{0}$, while its derivative, $d_{xx}=m\omega^{2}$, is nonzero, fulfilling the second and third requirements.

To illustrate how such a quadratic potential could arise from elementary particle physics consider an interaction resulting from the simultaneous exchange of a scalar and vector boson coupling to baryon number, whose potential energies naturally enter with opposite signs,
\begin{eqnarray}
\label{e:Vs}
V_{S}(\vec{r}) & = & -Gm_{i}m_{j}q_{i}q_{j}\frac{\xi_{S}}{r}e^{-r/\lambda_{S}}\text{,}\\
\label{e:Vv}
V_{V}(\vec{r}) & = & Gm_{i}m_{j}q_{i}q_{j}\frac{\xi_{V}}{r}e^{-r/\lambda_{V}}\text{,}
\end{eqnarray}
where $q_{k} = B_{k}/\mu_{k}$ is the baryon-number-dependent charge, and we have used the notation from Ref.~\cite{FischbachBook}. 

To carry out the integrations over the mass distributions, we note that the Earth serves as the dominant source for the EPF experiment, and one of two sources for the E-W experiment. If the Earth is modeled as a uniform sphere, with the experimental apparatus located a distance z above the surface, the integral over the volume of the Earth is given by
\begin{eqnarray}
V_{k} = \pm G m_{i}q_{i}\xi_{k}\int \mathrm{d}^{3}r^{\prime}\rho\left(\vec{r}^{\ \prime}\right)q_{\oplus}\left(\vec{r}^{\ \prime}\right)\frac{e^{-\left|\vec{r}-\vec{r}^{\ \prime}\right|/\lambda_{k}}}{\left|\vec{r}-\vec{r}^{\ \prime}\right|}\text{,}
\end{eqnarray}
where $V_{k}$ denotes $V_{S}$ or $V_{V}$ in Eqs.~(\ref{e:Vs}) and (\ref{e:Vv}). For a uniform sphere, $\rho_{\oplus}$ and $q_{\oplus}$ are constants in which case,
\begin{eqnarray}
\nonumber V_{k} &=& \pm G\rho_{\oplus} m_{i}q_{i}q_{\oplus}\xi_{k}\int_{0}^{2\pi} \mathrm{d}\phi^{\prime} \int_{0}^{R_{\oplus}} \mathrm{d}r^{\prime}{r^{\prime}}^{2}\\
&&\times\int_{0}^{\pi}\mathrm{d}\theta^{\prime}\sin\theta^{\prime}\frac{e^{-\sqrt{{r^{\prime}}^{2}+r^2-2rr^{\prime}\cos\theta^{\prime}}/\lambda_{k}}}{\sqrt{{r^{\prime}}^{2}+r^2-2rr^{\prime}\cos\theta^{\prime}}}\text{.}
\end{eqnarray}
Let $u = \sqrt{{r^{\prime}}^{2}+r^2-2rr^{\prime}\cos\theta^{\prime}}$, then $\mathrm{d}u = rr^{\prime}\sin\theta^{\prime}\mathrm{d}\theta^{\prime}/\sqrt{{r^{\prime}}^{2}+r^2-2rr^{\prime}\cos\theta^{\prime}}$, and integrating over $\phi^{\prime}$, yields
\begin{eqnarray}
\label{e:V_phiint}
V_{k} &=& \pm 2\pi \frac{G\rho_{\oplus} m_{i}\lambda_{k}}{r}q_{i}q_{\oplus}\xi_{k}\int_{0}^{R} \mathrm{d}r^{\prime}r^{\prime}\left(e^{-\left|r-r^{\prime}\right|/\lambda_{k}}-e^{-(r+r^{\prime})/\lambda_{k}}\right)\text{.}
\end{eqnarray}
Integrating over $r^{\prime}$ in Eq.~(\ref{e:V_phiint}) we then find,
\begin{eqnarray}
\label{e:V_Earth}
\nonumber V_{k} &=& \pm \frac{3}{2} \frac{GM_{\oplus} m_{i}}{R_{\oplus}+z}q_{i}q_{\oplus}\xi_{k} \left(\frac{\lambda_{k}}{R_{\oplus}}\right)^{2}e^{-z/\lambda_{k}}\\
 &&\times \left[\left(1-\frac{\lambda_{k}}{R_{\oplus}}\right)+\left(1+\frac{\lambda_{k}}{R_{\oplus}}\right)e^{-2R_{\oplus}/\lambda_{k}}\right]\text{.}
\end{eqnarray}
Since the apparatus is close to the surface ($z \ll R_{\oplus}$), and the force is of short range ($\lambda_{k} \ll R_{\oplus}$), Eq.~(\ref{e:V_Earth}) can be simplified to read,
\begin{eqnarray}
\label{e:Earthpot}
V_{k} &=& \pm \frac{3}{2} \frac{GM_{\oplus} m_{i}}{R_{\oplus}}q_{i}q_{\oplus}\xi_{k} \left(\frac{\lambda_{k}}{R_{\oplus}}\right)^{2}e^{-z/\lambda_{k}}\text{.}
\end{eqnarray}

The total potential and force are then given by,
\begin{eqnarray}
V_{\rm tot} &=& \frac{3}{2} \frac{GM_{\oplus} m_{i}}{R_{\oplus}}q_{i}q_{\oplus}\left[\xi_{V} \left(\frac{\lambda_{V}}{R_{\oplus}}\right)^{2}e^{-z/\lambda_{V}}-\xi_{S} \left(\frac{\lambda_{S}}{R_{\oplus}}\right)^{2}e^{-z/\lambda_{S}}\right]\text{,}\\
\nonumber \vec{F}_{\rm tot} &=& \frac{3}{2} \frac{GM_{\oplus} m_{i}}{R_{\oplus}^{2}}q_{i}q_{\oplus}\left[\xi_{V} \left(\frac{\lambda_{V}}{R_{\oplus}}\right)e^{-z/\lambda_{V}}-\xi_{S} \left(\frac{\lambda_{S}}{R_{\oplus}}\right)e^{-z/\lambda_{S}}\right]\hat{z}\\
&=& \frac{3}{2} g_{N} m_{i}q_{i}q_{\oplus}\left(\overline{\xi}_{V} e^{-z/\lambda_{V}}-\overline{\xi}_{S}e^{-z/\lambda_{S}}\right)\hat{z}.
\label{e:EarthForceZ}
\end{eqnarray}
In Eq.~(\ref{e:EarthForceZ}), we have introduced the gravitational acceleration due to Earth, $g_{N}$, and have defined $\overline{\xi_{V}}= \xi_{V} \left(\lambda_{V}/R_{\oplus}\right)$. To clarify the direction of the unit vector $\hat{z}$ in Eq.~(\ref{e:EarthForceZ}), we note that the integral over the Earth's volume used a standard spherical reference frame, in which our apparatus was located radially away from the center of the Earth at a distance of $R+z$. In our NED frame (Figure \ref{fi:EarthFrame}) this direction corresponds to $-\hat{D}$, and hence we can replace $\hat{z}$ with $-\hat{D}$ in Eq.~(\ref{e:EarthForceZ}).

The second source for the E-W experiment was a hill located North of the apparatus. For present purposes, this hill can be modeled as an infinite half plane with the apparatus located a distance z above the plane. We assume for simplicity that the hill has the same $q$ and $\rho$ as the Earth, and we express the mass element at $\mathrm{d}m_{j} = \rho_{\oplus}r^{\prime}\mathrm{d}r^{\prime}\mathrm{d}\phi^{\prime}\mathrm{d}z^{\prime}$ in cylindrical coordinates. Then,
\begin{eqnarray}
\nonumber V_{k} &=& \pm G\rho_{\oplus} m_{i}q_{i}q_{\oplus}\xi_{k}\int_{0}^{2\pi} \mathrm{d}\phi^{\prime} \int_{-\infty}^{0} \mathrm{d}z^{\prime}\\
&&\times\int_{0}^{\infty}\mathrm{d}r^{\prime}\, r^{\prime}\frac{e^{-\sqrt{{r^{\prime}}^{2}+(z-z^{\prime})^2}/\lambda_{k}}}{\sqrt{{r^{\prime}}^{2}+(z-z^{\prime})^2}}\text{.}
\end{eqnarray}
Substituting $u = \sqrt{{r^{\prime}}^{2}+(z-z^{\prime})^2}$, with $\mathrm{d}u = r^{\prime}\mathrm{d}r^{\prime}/\sqrt{{r^{\prime}}^{2}+(z-z^{\prime})^2}$, we find after integrating over $\phi^{\prime}$ 
\begin{eqnarray}
 V_{k} &=& \pm 2\pi G\rho_{\oplus} m_{i}q_{i}q_{\oplus}\xi_{k}\lambda_{k}\int_{-\infty}^{0} \mathrm{d}z^{\prime}\,e^{-\left|z-z^{\prime}\right|/\lambda_{k}}\text{.}
\end{eqnarray}
Since $z$ is positive and $z^{\prime}$ is always negative, we can drop the absolute value and write
\begin{eqnarray}
\nonumber V_{k} &=& \pm 2\pi G\rho_{\oplus} m_{i}q_{i}q_{\oplus}\xi_{k}\lambda_{k}e^{-z/\lambda_{k}}\int_{-\infty}^{0} \mathrm{d}z^{\prime}\,e^{z^{\prime}/\lambda_{k}}\\
&=& \pm \frac{3}{2} g_{N} m_{i}q_{i}q_{\oplus}\overline{\xi}_{k}\lambda_{k}e^{-z/\lambda_{k}}\text{,}
\label{e:Vk}
\end{eqnarray}
where we have introduced the same simplifications as before. It follows from Eqs.~(\ref{e:Earthpot}) and (\ref{e:Vk}) that the potential for the hill and the Earth have the same functional form. This makes sense since the assumed short range of the force implies that the apparatus only ``sees'' the contributions from the Earth in its immediate vicinity, and at that distance the Earth appears as a flat infinite plane. The force from the hill will then be given by Eq.~(\ref{e:EarthForceZ}), but we must again be careful about notation and vectors. In our model, the apparatus was located a distance $z$ above the semi-infinite plane.  However, in reality, the apparatus was located a distance $x$ to the South, so we must replace $z$ in Eq.~(\ref{e:EarthForceZ}) with $x$, and $\hat{z}$ with $-\hat{N}$. The forces from the Earth and the hill are then given by, 
\begin{eqnarray}
\label{e:hillpot}
\vec{F}_{\oplus} &=& -\frac{3}{2} g_{N} m_{i}q_{i}q_{\oplus}\left(\overline{\xi}_{V} e^{-z/\lambda_{V}}-\overline{\xi}_{S}e^{-z/\lambda_{S}}\right)\hat{D}\text{,}\\
\vec{F}_{\text{hill}}&=& -\frac{3}{2} g_{N} m_{i}q_{i}q_{\oplus}\left(\overline{\xi}_{V} e^{-x/\lambda_{V}}-\overline{\xi}_{S}e^{-x/\lambda_{S}}\right)\hat{N}\text{.}
\end{eqnarray}
The equilibrium positions of these functions can then be found by setting the total force equal to zero, from which it follows that,
\begin{eqnarray}
\label{e:zequil}
 z_{\rm equil} & = & \left(\frac{1}{\lambda_{V}}-\frac{1}{\lambda_{S}}\right)^{-1}\ln\left(\frac{\overline{\xi}_{V}}{\overline{\xi}_{S}}\right)\text{.} 
\end{eqnarray}
Since $z_{\rm equil}$ or $x_{\rm equil}$ must be positive, Eq.~(\ref{e:zequil}) gives rise to two possibilities:
\begin{subequations}
\label{e:positivity}
\begin{eqnarray}
\label{e:positivityA}
&&\frac{1}{\lambda_{V}}-\frac{1}{\lambda_{S}} > 0\text{, }\frac{\overline{\xi}_{V}}{\overline{\xi}_{S}} > 1,\\
\label{e:positivityB}
&&\frac{1}{\lambda_{V}}-\frac{1}{\lambda_{S}} < 0\text{, }\frac{\overline{\xi}_{V}}{\overline{\xi}_{S}} < 1.
\end{eqnarray} 
\end{subequations}

\subsubsection{Case 1: $\overline{\xi}_{V} > \overline{\xi}_{S}$, $\lambda_{S} > \lambda_{V}$}

The positivity conditions in Eq.~(\ref{e:positivityA}) can be rewritten as
\begin{subequations}
\begin{eqnarray}
\frac{\overline{\xi}_{S}}{\overline{\xi}_{V}} &<& 1\text{,}\\
\frac{\lambda_{V}}{\lambda_{S}} &<& 1\text{.}
\end{eqnarray}
\end{subequations}
We introduce parameters $\alpha_{1} = \overline{\xi}_{S}/\overline{\xi}_{V}$ and $\zeta_{1} = \lambda_{V}/\lambda_{S}$ which are bounded between zero and unity. The relevant phenomenological coefficients from Eq.~(\ref{e:fifthfor_q}) can then be expressed as,
\begin{subequations}
\begin{eqnarray}
f_{5x} &=& -\frac{3}{2} g_{N} q_{\oplus}\overline{\xi}_{V}
\left( e^{-x/\lambda_{V}}-\alpha_{1} e^{-x\zeta_{1}/\lambda_{V}}\right)\text{,}\\
f_{5z} &=& -\frac{3}{2} g_{N} q_{\oplus}\overline{\xi}_{V}
\left( e^{-z/\lambda_{V}}-\alpha_{1} e^{-z\zeta_{1}/\lambda_{V}}\right)\text{,}\\
d_{5zz} &=& \frac{3}{2} g_{N} q_{\oplus}\frac{\overline{\xi}_{V}
}{\lambda_{V}}\left( e^{-z/\lambda_{V}}-\alpha_{1}\zeta_{1} e^{-z\zeta_{1}/\lambda_{V}}\right)\text{.}
\end{eqnarray}
\end{subequations}
The E\"otv\"os parameters in the EPF and E-W experiments are then given by,
\begin{subequations}
\begin{eqnarray}
\label{e:EPFCase1}
\nonumber \frac{\eta_{\rm EPF}}{\frac{3}{2}\Delta q q_{\oplus}\overline{\xi}_{V}} & \equiv & \overline{\eta}_{\rm EPF} =
-s_{\beta}\left[ e^{-z_{\rm EPF}/\lambda_{V}}-\alpha_{1} e^{-z_{\rm EPF}\zeta_{1}/\lambda_{V}}\right]\\
& & + 
s_{\beta}\frac{h}{\lambda_{V}}\left[ e^{-z_{\rm EPF}/\lambda_{V}}-\alpha_{1} \zeta_{1} e^{-z_{\rm EPF}\zeta_{1}/\lambda_{V}}\right]\text{,}\\
\nonumber \frac{\eta_{\rm EW}}{\frac{3}{2}\Delta q q_{\oplus}\overline{\xi}_{V}} & \equiv & \overline{\eta}_{\rm EW} = \sqrt{2}\biggl|-\left( e^{-x_{\rm EW}/\lambda_{V}}-\alpha_{1} e^{-x_{\rm EW}\zeta_{1}/\lambda_{V}}\right)\\
&& - s_{\beta}\left( e^{-z_{\rm EW}/\lambda_{V}}-\alpha_{1} e^{-z_{\rm EW}\zeta_{1}/\lambda_{V}}\right)\biggr|\text{.}
\label{e:EWCase1}
\end{eqnarray}
\end{subequations}
We note that since $\overline{\xi}_{V}$ simply scales the magnitude of any effect, it is convenient to move it to the left side of the above equations. This leaves only three parameters which govern the magnitude of the effect: $\lambda_{V}$, $\alpha_{1}$, and $\zeta_{1}$. 

\begin{figure}[t]
  \centering
  \includegraphics[width=\columnwidth]{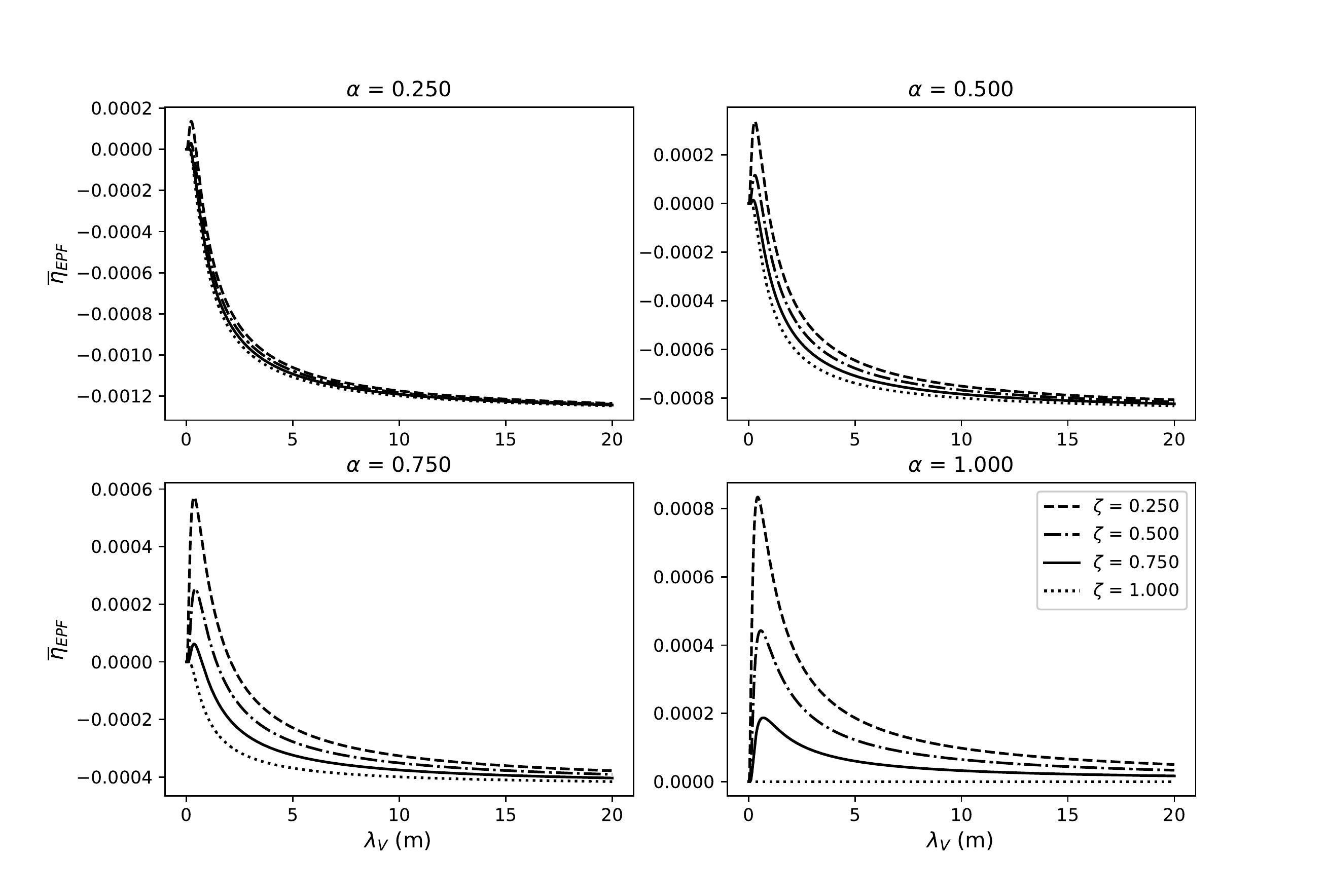}
  \caption{Reduced E\"otv\"os parameter for the E\"otv\"os experiment for various values of $\alpha_{1}$, $\zeta_{1}$, and $\lambda_{V}$.}
  \label{fi:EPFPlot1}
\end{figure}

\begin{figure}[t]
  \centering
  \includegraphics[width=\columnwidth]{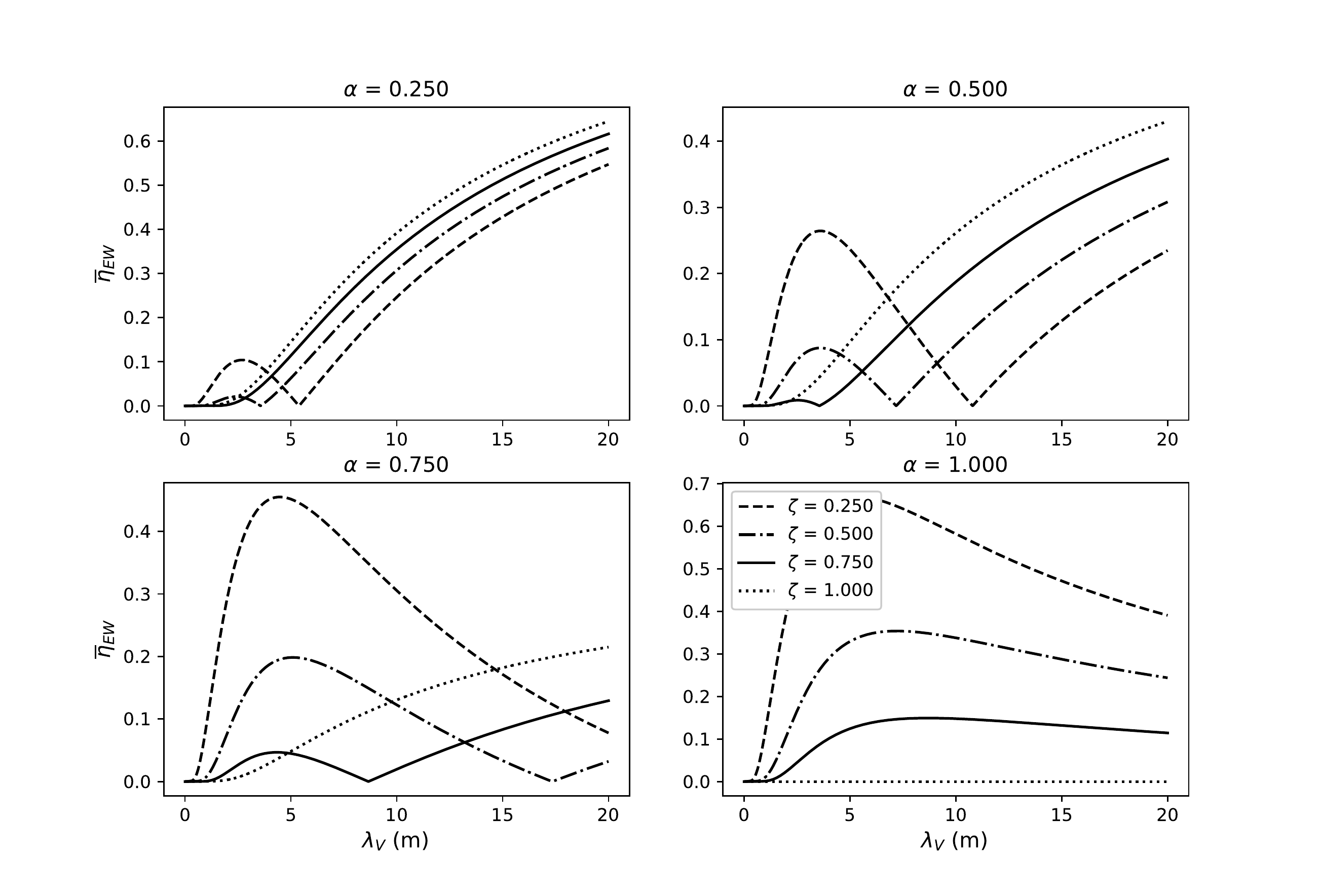}
  \caption{Reduced E\"otv\"os parameter for the E\"ot-Wash experiment for various values of $\alpha_{1}$, $\zeta_{1}$, and $\lambda_{V}$.}
  \label{fi:EWPlot1}
\end{figure}

It is instructive to plot Eqs.~(\ref{e:EPFCase1}) and (\ref{e:EWCase1}) for various values of $\alpha_{1}$, $\zeta_{1}$, and $\lambda_{V}$. We use  $s_{\beta} = 1.73\times 10^{-3}$, $h = 21.2$~cm\cite{EPF}, $x_{\rm EW}=10$~m, $z_{\rm EPF}=z_{\rm EW}=1$~m. The plots are shown in Fig. \ref{fi:EPFPlot1} for EPF and Fig. \ref{fi:EWPlot1} for E-W. We note that in Fig. \ref{fi:EPFPlot1}, the results are negative for small $\alpha_{1}$ values and eventually become positive as $\alpha_{1}$ increases. As $\alpha_{1}$ determines the relative strengths of the two Yukawas, we would expect $\alpha_{1}$ to have a large effect on the sign of the response. Fig. \ref{fi:EWPlot1} starts positive and would eventually become negative if not for the absolute value in Eq.~(\ref{e:EWCase1}); instead it grows positive. The zeros of $\eta_{\rm EW}$ occur at higher $\lambda_{V}$ for increasing $\alpha_{1}$ and fixed $\zeta_{1}$ and occur at lower $\lambda_{V}$ for increasing $\zeta_{1}$ at fixed $\alpha_{1}$. 
We notice in Fig. \ref{fi:EWPlot1} that for any $\alpha_{1}$ and $\zeta_{1}$, there exists a value of $\lambda_{V}$where the net signal is zero, which was one of the requirements for this force as discussed in Sect~\ref{sec:SV_Model}. In fact, these values can be determined analytically, by setting $\overline{\eta}_{\rm EW}=0$ in Eq.~(\ref{e:EWCase1}): We find
\begin{equation}
\alpha_{1} = \frac{e^{-x_{\rm EW}/\lambda_{V}}+ s_{\beta}e^{-z_{\rm EW}/\lambda_{V}}}{e^{-x_{\rm EW}\zeta_{1}/\lambda_{V}}+ s_{\beta}e^{-z_{\rm EW}\zeta_{1}/\lambda_{V}}}\text{.}
\end{equation}

\begin{figure}[t]
  \centering
  \includegraphics[width=\columnwidth]{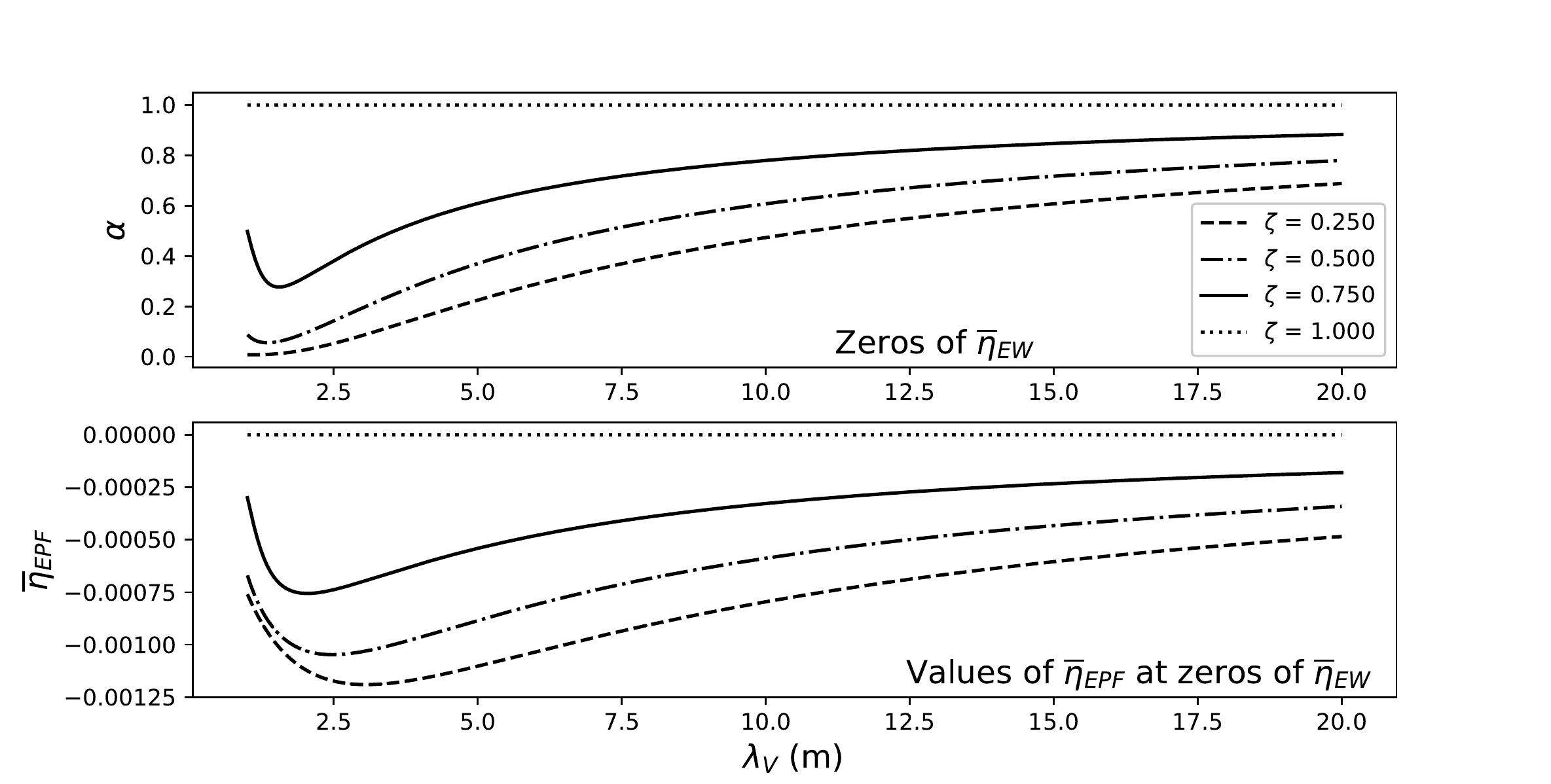}
  \caption{The top plot shows zeros of $\overline{\eta}_{\rm EW}$. The bottom plot shows the values of $\overline{\eta}_{\rm EPF}$ at the zeros of $\overline{\eta}_{\rm EW}$.}
  \label{fi:Image_alpha1}
\end{figure}

\noindent The top plot of Fig. \ref{fi:Image_alpha1} shows the values of $\alpha_{1}$ which give $\overline{\eta}_{\rm EW}=0$ for various values of $\zeta_{1}$ and $\lambda_{V}$. Interestingly, $\alpha_{1}$ remains less than or equal to one, as required by our initial definition. The bottom plot shows the value of $\overline{\eta}_{\rm EPF}$ for each triplet $(\lambda_{V},\zeta_{1},\alpha_{1})$. As before the answer is negative, and its magnitude peaks near $\lambda_{V} \approx (2-3)$ m for the entire range of $\zeta_{1}$, with the actual value decreasing as $\zeta_{1}$ increases.

\subsubsection{Case 2: $\overline{\xi}_{S} > \overline{\xi}_{V}$, $\lambda_{V} > \lambda_{S}$}

We now write the positivity conditions as
\begin{subequations}
\begin{eqnarray}
\frac{\overline{\xi}_{V}}{\overline{\xi}_{S}} &\le& 1\text{,}\\
\frac{\lambda_{S}}{\lambda_{V}} &\le& 1\text{.}
\end{eqnarray}
\end{subequations}
If we introduce parameters $\alpha_{2} = \overline{\xi}_{V}/\overline{\xi}_{S}$ and $\zeta_{s} = \lambda_{S}\lambda_{V}$ which are bounded between zero and unity, the relevant phenomenological coefficients from Eq.~(\ref{e:fifthfor_q}) can be written as
\begin{subequations}
\begin{eqnarray}
f_{5x} &=& -\frac{3}{2} g_{N} q_{\oplus}\overline{\xi}_{S}
\left( \alpha_{2}e^{-x\zeta_{2}/\lambda_{s}}- e^{-x/\lambda_{S}}\right)\text{,}\\
f_{5z} &=& -\frac{3}{2} g_{N} q_{\oplus}\overline{\xi}_{S}
\left(\alpha_{2}e^{-z\zeta_{2}/\lambda_{S}}-e^{-z/\lambda_{S}}\right)\text{,}\\
d_{5zz} &=& \frac{3}{2} g_{N} q_{\oplus}\frac{\overline{\xi}_{S}
}{\lambda_{S}}\left(\alpha_{2}\zeta_{2} e^{-z\zeta_{2}/\lambda_{S}}-e^{-z/\lambda_{S}}\right)\text{.}
\end{eqnarray}
\end{subequations}

\begin{figure}[t]
  \centering
  \includegraphics[width=\columnwidth]{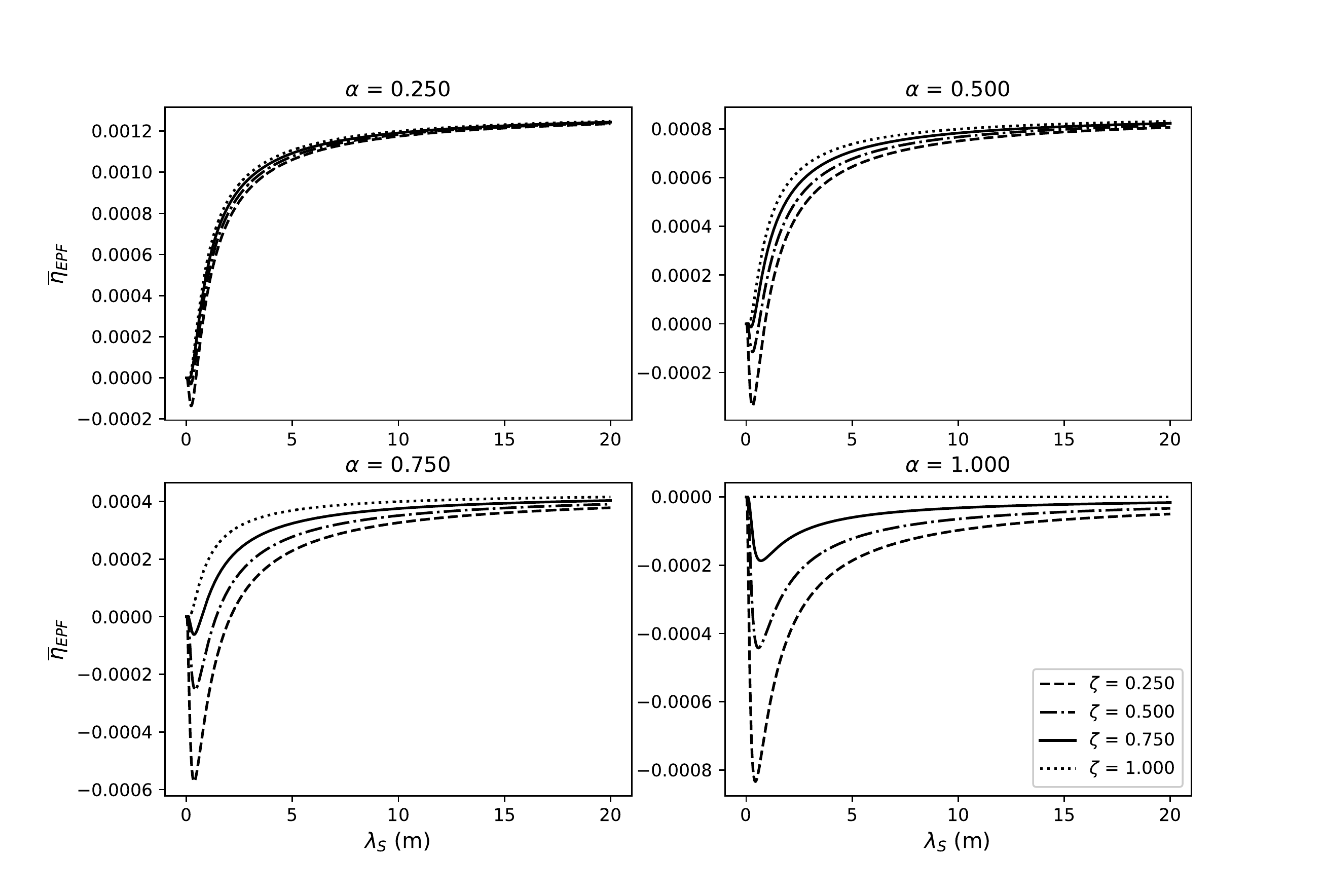}
  \caption{Reduced E\"otv\"os parameter for the E\"otv\"os experiment for various values of $\alpha_{2}$, $\zeta_{2}$, and $\lambda_{S}$.}
  \label{fi:EPFPlot2}
\end{figure}

\begin{figure}[t]
  \centering
  \includegraphics[width=\columnwidth]{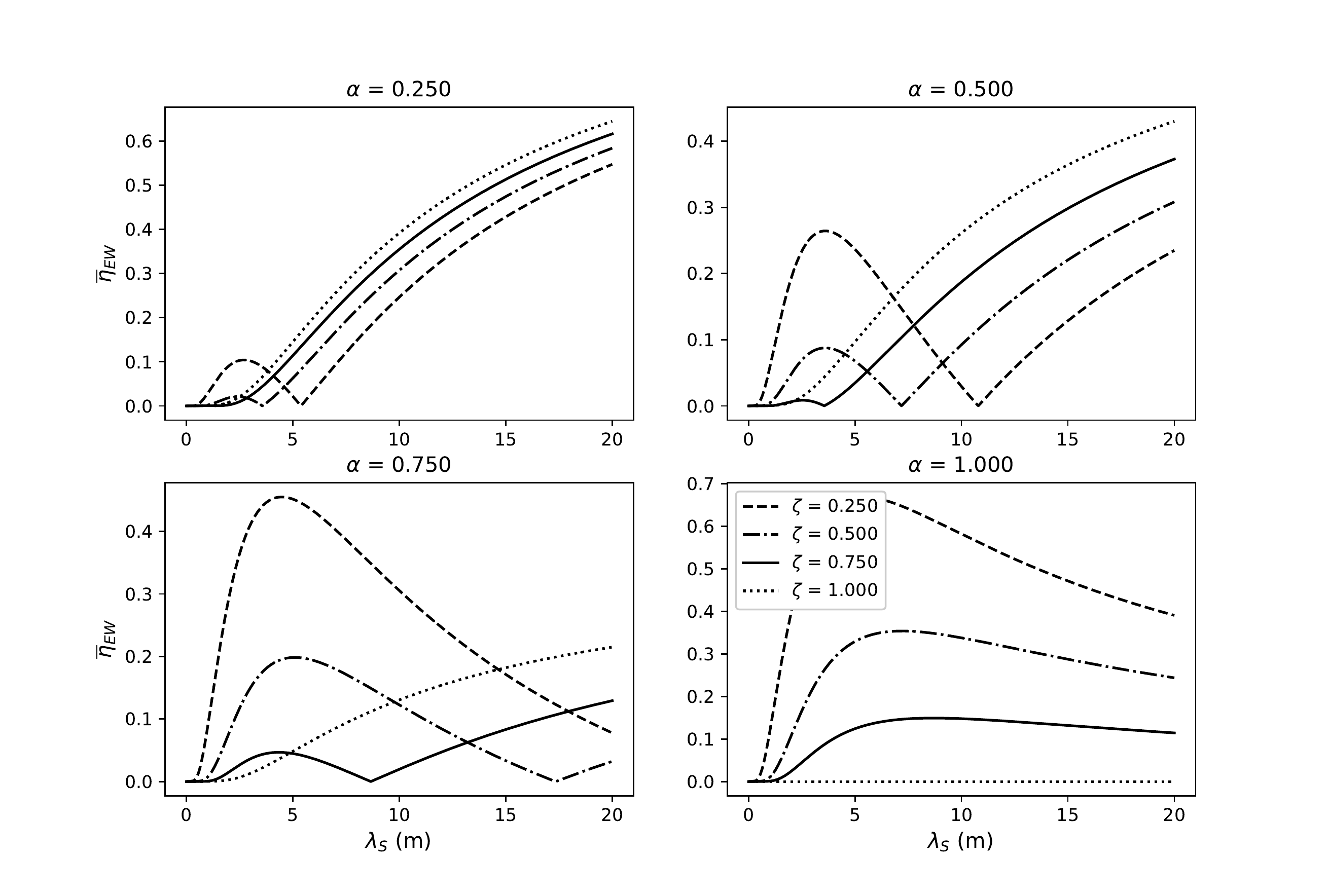}
  \caption{Reduced E\"otv\"os parameter for the E\"ot-Wash experiment for various values of $\alpha_{2}$, $\zeta_{2}$, and $\lambda_{S}$.}
  \label{fi:EWPlot2}
\end{figure}

The E\"otv\"os parameters in the EPF and E-W experiments are then given by
\begin{subequations}
\begin{eqnarray}
\nonumber \frac{\eta_{\rm EPF}}{\frac{3}{2}\Delta q q_{\oplus}\overline{\xi}_{S}} & = & \overline{\eta}_{\rm EPF} = 
-s_{\beta}\left[ \alpha_{2}e^{-z_{\rm EPF}\zeta_{2}/\lambda_{S}}-e^{-z_{\rm EPF}/\lambda_{S}}\right]\\
& & + 
s_{\beta}\frac{h}{\lambda_{S}}\left[ \alpha_{2}\zeta_{2} e^{-z_{\rm EPF}\zeta_{2}/\lambda_{S}}-e^{-z_{\rm EPF}/\lambda_{S}}\right]\text{,}\\
\label{e:EPFCase2}
\nonumber \frac{\eta_{\rm EW}}{\frac{3}{2}\Delta q q_{\oplus}\overline{\xi}_{S}} & = & \overline{\eta}_{\rm EW} = \sqrt{2}\biggl|-\left( \alpha_{2}e^{-x_{\rm EW}\zeta_{2}/\lambda_{s}}- e^{-x_{\rm EW}/\lambda_{S}}\right)\\
&& - s_{\beta}\left( \alpha_{2}e^{-z_{\rm EW}\zeta_{2}/\lambda_{S}}-e^{-z_{\rm EW}\/\lambda_{S}}\right)\biggr|\text{.}
\label{e:EWCase2}
\end{eqnarray}
\end{subequations}
We notice that $\overline{\eta}_{\rm EW}$ will be unchanged from Eq.~(\ref{e:EWCase1}) due to the absolute value, while $\overline{\eta}_{\rm EPF}$ will change sign from Eq.~(\ref{e:EPFCase1}). These conclusions are supported by Figs. \ref{fi:EPFPlot2} and \ref{fi:EWPlot2}. 

\begin{figure}[b]
  \centering
  \includegraphics[width=\columnwidth]{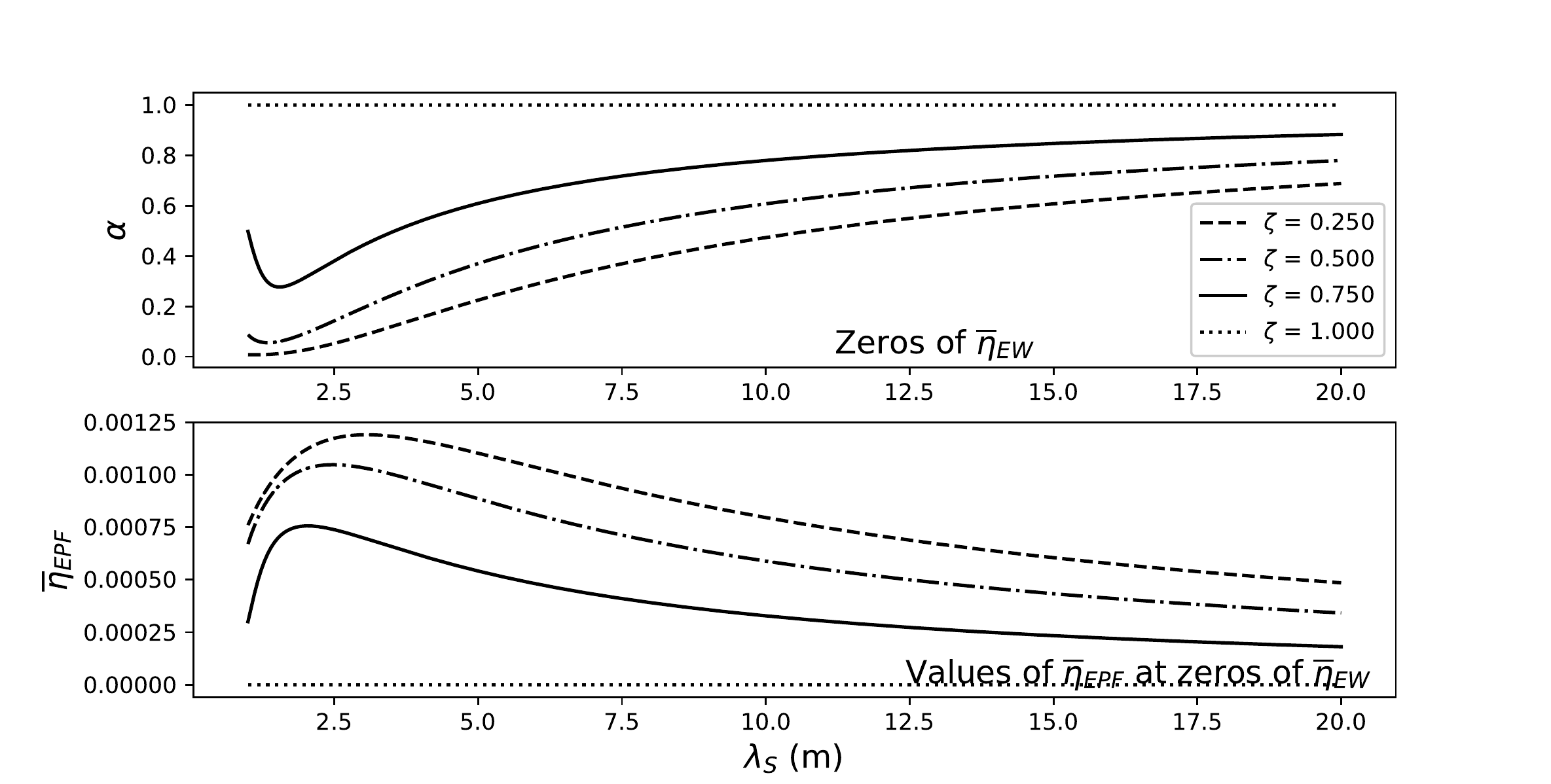}
  \caption{The top plot shows zeros of $\eta_{\rm EW}$ for Case 2. The bottom plot shows value of $\overline{\eta}_{\rm EPF}$ at the zeros of $\overline{\eta}_{\rm EW}$.}
  \label{fi:Image_alpha2}
\end{figure}

We can once again search for the zeros of $\overline{\eta}_{\rm EW}$, which are given by Eq.~(\ref{e:zeros2}). The results are shown in Fig. \ref{fi:Image_alpha2} and we obtain the same result as in Fig. \ref{fi:Image_alpha1}, but with $\overline{\eta}_{\rm EPF}$ having the opposite sign as seen in  Fig. \ref{fi:Image_alpha1}
\begin{equation}
\alpha_{2} = \frac{e^{-x_{\rm EW}/\lambda_{S}}+ s_{\beta}e^{-z_{\rm EW}/\lambda_{S}}}{e^{-x_{\rm EW}\zeta_{2}/\lambda_{S}}+ s_{\beta}e^{-z_{\rm EW}\zeta_{2}/\lambda_{S}}}\text{.}
\label{e:zeros2}
\end{equation}

\subsubsection{Numerical Analysis}

We conclude this section with an estimate of $\xi_{S}$. We choose Case 2 since that will give us the positive $\overline{\eta}_{\rm EW}$. From Figure \ref{fi:Image_alpha2}, we select, $\zeta_{2} = 0.5$, $\lambda_{S}=2.5$m, which gives $\alpha_{2}=0.2$ and $\overline{\eta}_{\rm EPF}=10^{-3}$. We also note that $\eta_{\rm EPF} = \Delta\kappa s_{\beta} \approx 10^{-9}$\cite{Fischbach1986}. We can then write $\xi_{S}$ as 
\begin{subequations}
\begin{eqnarray}
\xi_{S} & = & \frac{R_{\oplus}}{\lambda_{S}} \overline{\xi}_{S} = \frac{\eta_{\rm EPF}}{\frac{3}{2}\Delta q q_{\oplus}\overline{\eta}_{\rm EPF}}\frac{R_{\oplus}}{\lambda_{S}},\\
 \nonumber \xi_{V} & = & \frac{R_{\oplus}}{\lambda_{V}} \overline{\xi}_{V} =\frac{R_{\oplus}}{\lambda_{V}} \alpha_{2}\overline{\xi}_{S}=\frac{\lambda_{S}}{\lambda_{V}} \alpha_{2}\xi_{S}= \alpha_{2}\zeta_{2}\xi_{S}\\
 & = & \alpha_{2}\zeta_{2}\frac{\eta_{\rm EPF}}{\frac{3}{2}\Delta q q_{\oplus}\overline{\eta}_{\rm EPF}}\frac{R_{\oplus}}{\lambda_{S}}.
\end{eqnarray}
\end{subequations}
From Ref.~\cite{Fischbach1986}, $\Delta q \approx 10^{-3}$ and $q_{\oplus} \approx 1$. Using $R_{\oplus} = 6\times 10^{6}$m, we find $\xi_{S} = 1600$ and $\xi_{V} = 160$. Solving for the coupling constant $f_{S,V}^{2} = \xi_{S,V} Gm_{H}^2$, we then find 
\begin{subequations}
\begin{eqnarray}
f^{2}_{S} &=& 2.98\times 10^{-55}~{\rm J \cdot m} = 9.42\times 10^{-30} \hbar c,\\
f^{2}_{V} &=& 2.98\times 10^{-56}~{\rm J \cdot m}= 9.42\times 10^{-31} \hbar c.
\end{eqnarray}
\end{subequations}
Thus, within this toy model using the assumed set of parameters, it is possible to resolve the \Eotvos\ paradox by assuming the existence of a 5th force that would not have been observed in E-W experiment.   However, modifications to the E-W setup, such as changing its position, would enable it to observe this scalar-vector 5th force.  More generally, however, these results demonstrate the possibility that only one of these experiments would have been sensitive to a new force.

\subsection{Forces with a Nonzero Curl}
\label{section:nonzero curl}

As discussed in the Introduction, many experiments have excluded 5th force models based on simple Yukawa potentials of the form given by Eq.~(\ref{Yukawa}).   Following Eq.~(\ref{e:T_tot}), we noted that the last set of terms (denoted by NZC) have a nonzero curl, and hence cannot describe the interaction of test samples with a force arising from the gradient of a potential.  However, a  ``magnetic'' 5th force analogous to the usual magnetic or gravimagnetic forces would have this character.  Recently Berry and Shukla have investigated so-called ``curl force'' dynamics \cite{Berry2016,Berry2012}.  In addition, nonzero curl forces also arise in fluid dynamics, which in our case might be due to a ``dark medium'' (e.g., dark matter or dark energy).  An example would be a force arising from a ``dark matter wind'' as the Earth moves through the galaxy's dark matter halo \cite{Dvorkin2014}.

Another example of a nonzero curl force can arise from the interactions of test samples due to the motion of the Earth through an external  dark medium in the presence of a gravitational interaction.   As we will show elsewhere, there can arise from the medium a 5th force contribution whose direction is determined by both the local gravitational force $\vec{F}_{g}$, and the velocity $\vec{v}$ of the Earth through the medium. The new feature is that this direction is in general different from both  $\hat{F}_{g}$ and $\hat{v}$, and hence could be misidentified as a a spurious external perturbation. This suggests a new class of experiments to search for a possible 5th force, as we will discuss in detail elsewhere.

\section{Discussion}
\label{discussion}

As noted in Sec.~\ref{sec:Introduction}, there is compelling evidence to support the claims that the EPF experiment \cite{EPF} was done correctly, and, additionally, that the reanalysis of the EPF data reported in \cite{Fischbach1986} was also correct, leading to the suggestion of a 5th force.  The failure to date of many experiments to detect the 5th force proposed in \cite{Fischbach1986} is thus puzzling.  Our objective in the present paper has been to demonstrate that there may exist a broader class of theories beyond that proposed in \cite{Fischbach1986} which could also account for the EPF data in \cite{EPF}, while at the same time avoiding detection to date. A  common feature of these alternative theories is that they could give rise to signals coming from unexpected directions, which could have been misidentified as spurious perturbations.  As an example, we demonstrated that the differences in the designs of the EPF and E\"{o}t-Wash experiment \cite{Eotwash1987} makes them sensitive to different 5th force signals. In future work we will use the formalism presented here to address the question of whether a specific 5th force interaction exists which would have shown up in the EPF experiment, but not in more recent 5th force searches.


\end{document}